\documentclass[ 
twocolumn,
showpacs,preprintnumbers,
bibnotes,
 amsmath,amssymb,
 aps,
 pra,
superscriptaddress,
longbibliography,
]{revtex4-1}

\usepackage[version=3]{mhchem} % Formula subscripts using \ce{}
\usepackage[dvipsnames]{xcolor}
\usepackage{textcomp}
\usepackage{graphicx}
\usepackage{amsmath}
\usepackage{fixmath}
\usepackage{amsfonts}
\usepackage{amssymb}
\usepackage{braket}
\usepackage{subfigure}
\usepackage{footnote}
\usepackage{tablefootnote}
\usepackage[normalem]{ulem}
\definecolor{pacificb}{HTML}{1CA9C9}

\begin{document}

\title{All-optical magnetization control in CrI$_3$ monolayers: a microscopic theory}% Force line breaks 

\author{A. Kudlis}
\affiliation{Abrikosov Center for Theoretical Physics, MIPT, Dolgoprudny, Moscow Region 141701, Russia}
\email{kudlis.a@mipt.ru}

\author{M. Kazemi}
\affiliation{Science Institute, University of Iceland, Dunhagi-3, IS-107 Reykjavik, Iceland}
\email{mak99@hi.is}

\author{Y. Zhumagulov}
\affiliation{University of Regensburg, Regensburg 93040, Germany}
\email{yaroslav.zhumagulov@gmail.com}

\author{H. Schrautzer}
\affiliation{Science Institute, University of Iceland, Dunhagi-3, IS-107 Reykjavik, Iceland}
\email{hes93@hi.is}

\author{A. I. Chernov}
\affiliation{Russian Quantum Center, Skolkovo, Moscow 143025, Russia}
\affiliation{Center for Photonics and 2D Materials, Moscow Institute of Physics and Technology (National Research University), Dolgoprudny 141700, Russia}
\email{ a.chernov@rqc.ru}

\author{P. F. Bessarab}
\affiliation{Science Institute, University of Iceland, Dunhagi-3, IS-107 Reykjavik, Iceland}
\affiliation{Department of Physics and Electrical Engineering, Linnaeus University, SE-39231 Kalmar, Sweden}
\email{pavel.bessarab@lnu.se}

\author{I. V. Iorsh}
\affiliation{Abrikosov Center for Theoretical Physics, MIPT, Dolgoprudny, Moscow Region 141701, Russia}
\email{i.iorsh@metalab.ifmo.ru}

\author{I. A. Shelykh}
\affiliation{Science Institute, University of Iceland, Dunhagi-3, IS-107 Reykjavik, Iceland}
\affiliation{Abrikosov Center for Theoretical Physics, MIPT, Dolgoprudny, Moscow Region 141701, Russia}
\email{shelykh@hi.is}

\date{\today}

\begin{abstract}
Bright excitons in ferromagnetic monolayers CrI$_3$ efficiently interact with lattice magnetization, which makes all-optical resonant magnetization control possible in this material. Using the combination of ab-initio simulations within Bethe-Salpeter approach, semiconductor Bloch equations, and Landau-Lifshitz equations, we construct a microscopic theory of this effect. By solving numerically the resulting set of coupled equations describing the dynamics of atomic spins and spins of the excitons, we demonstrate the possibility of a tunable control of macroscopic magnetization of a sample.
\end{abstract}

\maketitle
\section{Introduction}

Efficient control of the properties of layered structures is an important problem in both fundamental and applied research. In particular, a great demand exists for the development of optimal methods for control of magnetic characteristics of materials. This is not surprising, given the ever-increasing requirements for data-recording capacities of magnetic memory elements. The most important of them are compactness, energy efficiency, and recording speed. Regarding the latter, all-optical methods for the magnetic order manipulation look very promising as compared to traditional approaches based on the application of external magnetic field, as device operating frequencies can be enhanced by several orders of magnitude. 

To date, there already exists a number of theoretical \cite{PhysRevB.94.184406,PhysRevB.96.014409} and experimental works confirming the possibility of all-optical magnetization switching (AOMS) in a variety of magnetic compounds, including GdFeCo~\cite{Ignatyeva2019,Alives2020,Stanciu2020,Davies2020,Igrashi2020} and TbFeCo~\cite{Lu2018} ferrimagnetic alloys, as well as in Pt/Co or Co/Gd  multilayers~\cite{PhysRevB.94.064412,vanHees2020}. 

\begin{figure*}
    \centering
    \includegraphics[width=1.0\textwidth]{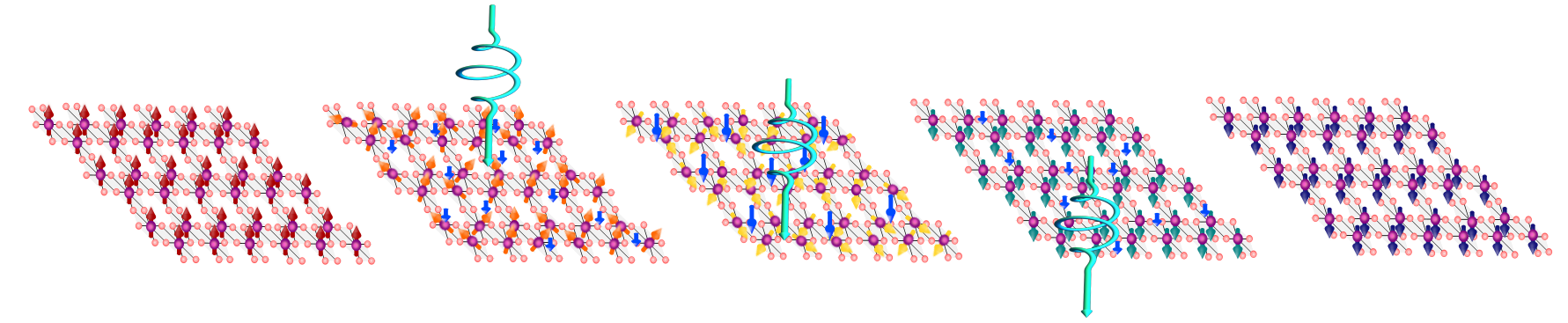}
    \caption{Schematic representation of the optical magnetization switching in CrI$_3$ monolayer. Here we demonstrate the system evolution in time (left to right). The monolayer is irradiated by circularly polarized light. Optical pumping leads to the formation of excitons carrying magnetic moment defined by the pump helicity which interact with the lattice magnetization of the sample.}
    \label{fig:img1}
\end{figure*}

Among all candidates where such a reorientation of magnetization is possible, materials combining ferromagnetic ordering with the presence of robust bright excitons are of  particular interest \cite{zhang2022all}. The examples of such materials are chromium trichalides, such as CrCl$_3$, CrBr$_3$ or CrI$_3$ \cite{huang2018electrical}. In the present work, we take CrI$_3$ as an example, but the reported results should remain qualitatively the same for other members of the family. In the material we consider,  the Cr$^{3+}$ ions are arranged on the honeycomb lattice vertices surrounded by non-magnetic I$^-$ ions  \cite{mcguire2015coupling}.  Being a 2D Ising ferromagnet, this material demonstrates robust optical excitonic response, with record high values of excitonic binding energies and oscillator strengths \cite{Wu2019}, exceeding even the values reported for transition metal dichalcogenides \cite{Chernikov2014,Splendiani2010,Steinleitner2017,Wang2018}.

The combination of such unique properties allowed some of us to propose that chromium trichalides were suitable candidates for polarization-sensitive resonant optical magnetization switching~\cite{PhysRevB.104.L020412}, which was later on confirmed experimentally~\cite{zhang2022all}. The process of magnetic reorientation is connected with the transfer of angular momentum from excitons (electron-hole pairs) to the quasi-localized \textit{d}-electrons of the Cr atoms, and corresponding phenomenological theory was developed in Ref. \onlinecite{PhysRevB.104.L020412}. However, the full microscopic theory of the effect is still absent.  

In this article, we make an attempt to construct such microscopic theory. We apply the well-established atomistic spin dynamics (ASD) formalism as basis of our work and couple it with the equations for the exciton dynamics by adding the terms describing the interaction between the spins of the excitons and the magnetic lattice. Numerical solution of the resulting set of equations allows us to analyze in detail the dynamics of the magnetization switching in real space and time. 

\begin{figure}[t!]
    \centering
    \includegraphics[width = 1.0\linewidth]
    {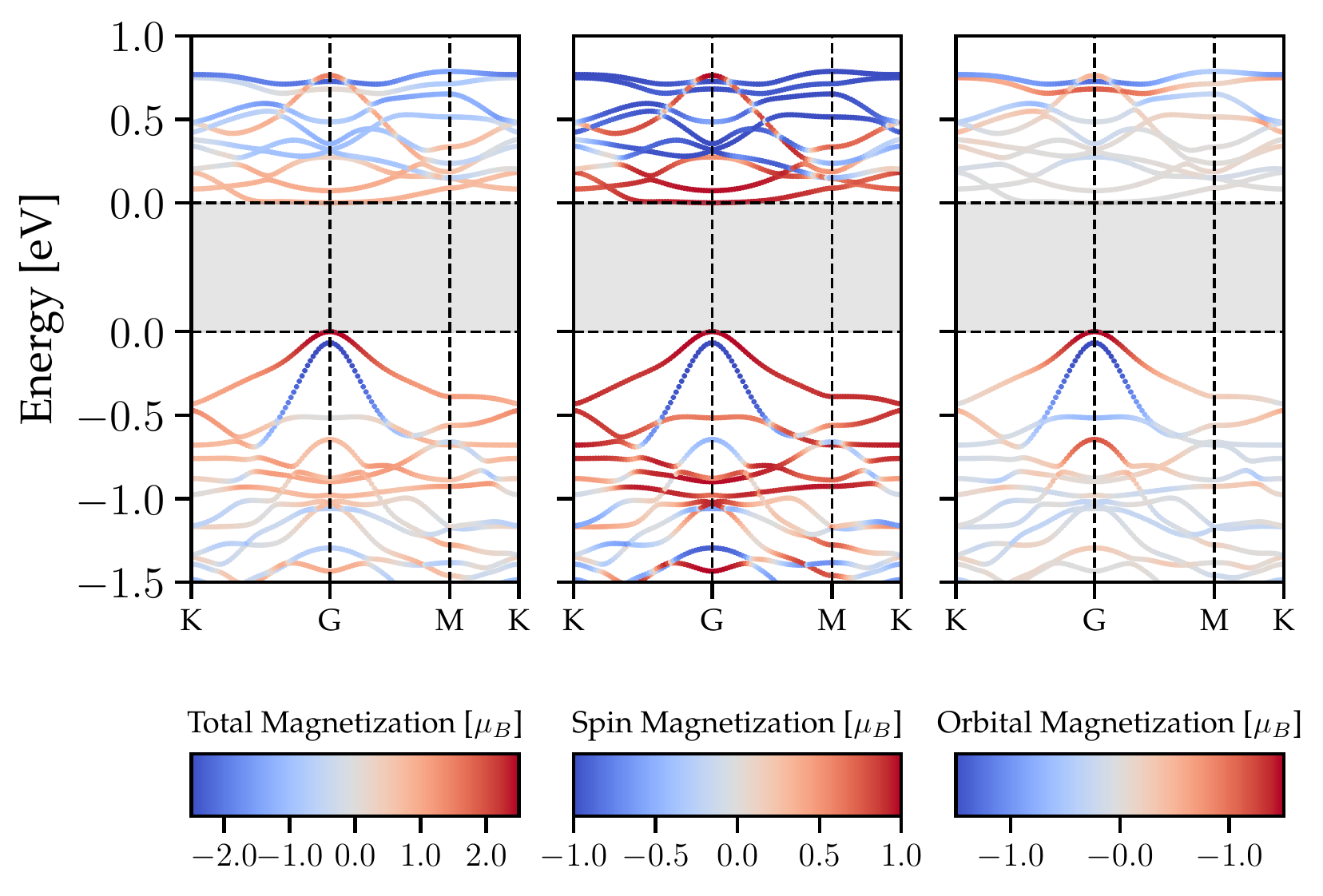}
    \caption{DFT band structure of excitons in  CrI$_3$ monolayer with lattice constant  $6.69$ {\AA}. The color code shows total, spin, and orbital  magnetization in the left, middle and right panel, respectively. The bandgap energy E=$2.59$~eV is indicated by gray shading. According to the first Brillouin zone grid, we have selected 16 valence bands and 14 conduction bands for the exciton basis.}
    \label{fig:M_band}
\end{figure}

The article is organized as follows. In Sec. II, we present calculations of the excitonic parameters in CrI$_3$, and then present the model Hamiltonian for coupled systems of excitons and lattice spins in Sec. III. Sec. IV presents the dynamic equations, and Sec. V contains the main results of the work, including the dependence of the switching properties on the parameters of the incident light beam. Sec. VI summarizes the results of the work.

\section{DFT/BSE calculations of exciton parameters for CrI$_{3}$}\label{subsec:dft}
\begin{figure}[t!]
    \centering
    \includegraphics[width = 1.0\linewidth]
    {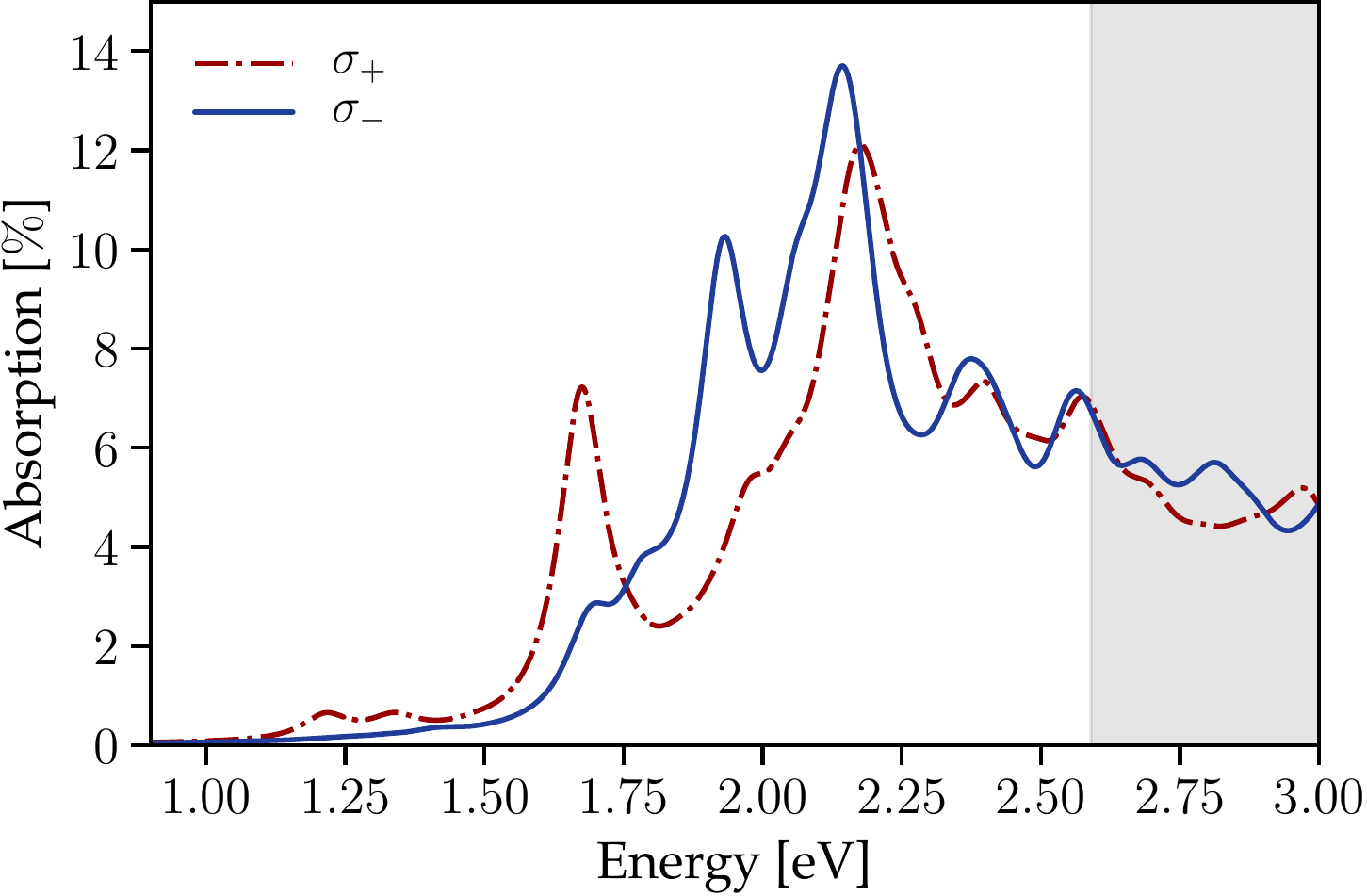}
    \caption{ Absorption spectrum of CrI$_3$ sample for two circular polarizations. The solid line and the dashed line correspond to the $\sigma_-$ and $\sigma_+$ polarizations, respectively. The bandgap energy E=$2.59$~eV is indicated by gray shading.}
    \label{fig:M_spec}
\end{figure}

We investigate the electronic structure of ferromagnetic CrI$_3$ monolayers via first-principles calculations employing density functional theory (DFT). The computations were executed using the \textsf{GPAW} package \cite{Mortensen_2005, Enkovaara_2010}. We use the cut-off energy of 600 eV for the plane-wave basis set and the LDA exchange-correlation functional incorporating spin-orbit effects \cite{PhysRevB.94.235106}. To mitigate interaction between periodic images, 16 {\AA} vacuum was used in the supercell.

Lattice constant for the CrI$_3$ ferromagnetic monolayer was determined via crystal lattice relaxation procedures, resulting in the value of $6.69$ {\AA}. The force convergence criteria were set at 1 meV/{\AA} per atom. To address the DFT bandgap issue, a scissor operator was employed to adjust the bandgap value to the experimental value of $2.59$ eV \cite{Wu2019}.

The exciton spectrum was acquired utilizing the \textsf{GPAW} implementation of the Bethe-Salpeter equation \cite{PhysRevB.62.4927, PhysRevB.83.245122, PhysRevB.88.245309, PhysRevLett.127.166402}. Screened Coulomb potential expressions were derived with the dielectric cutoff of 50 eV, 200 electronic bands, and a 2D truncated Coulomb potential \cite{PhysRevB.88.245309}. The exciton basis was configured with 16 valence bands and 14 conduction bands on a 6$\times$6 grid of the first Brillouin zone.

In Fig. \ref{fig:M_band}, we demonstrate the band structure of the CrI$_3$ monolayer with color bars for the orbital, spin and total magnetization which are shown below the main figures. The bandgap of $2.59$~eV is highlighted by the gray area. The analysis of the polarization-resolved absorption spectrum of the CrI$_3$ monolayer is shown in Fig. \ref{fig:M_spec}. As one can see, there is a remarkable difference in the absorption of the $\sigma_+$ and $\sigma_-$ components.  The prominent absorption peaks in both absorption profiles are related to the excitonic transitions. Naturally, if the magnetization of the sample is inverted, the absorption curves corresponding to opposite polarizations interchange. This forms the basis of the magnetization switching mechanism: for a given pump frequency and polarization, the magnetic sublattice tends to orient its magnetization so as to maximize the  absorption~\cite{PhysRevB.104.L020412}.

\section{The model Hamiltonian} \label{subsec:magnetization}

The total energy $E$ of the system includes three terms:
\begin{equation}
    E=E_{\text{m}}+E_{\text{exc}}+E_{\text{s}},
\end{equation}
describing the contributions from the magnetic subsystem, excitonic subsystem, and the interaction between them. 
The magnetic structure of the CrI$_3$ monolayer 
is described within 
the model of classical spin vectors localized on sites of the honeycomb lattice of the Cr atoms. The corresponding energy 
is given by:
\begin{figure}
    \centering
    \includegraphics[width = 1.0\linewidth]{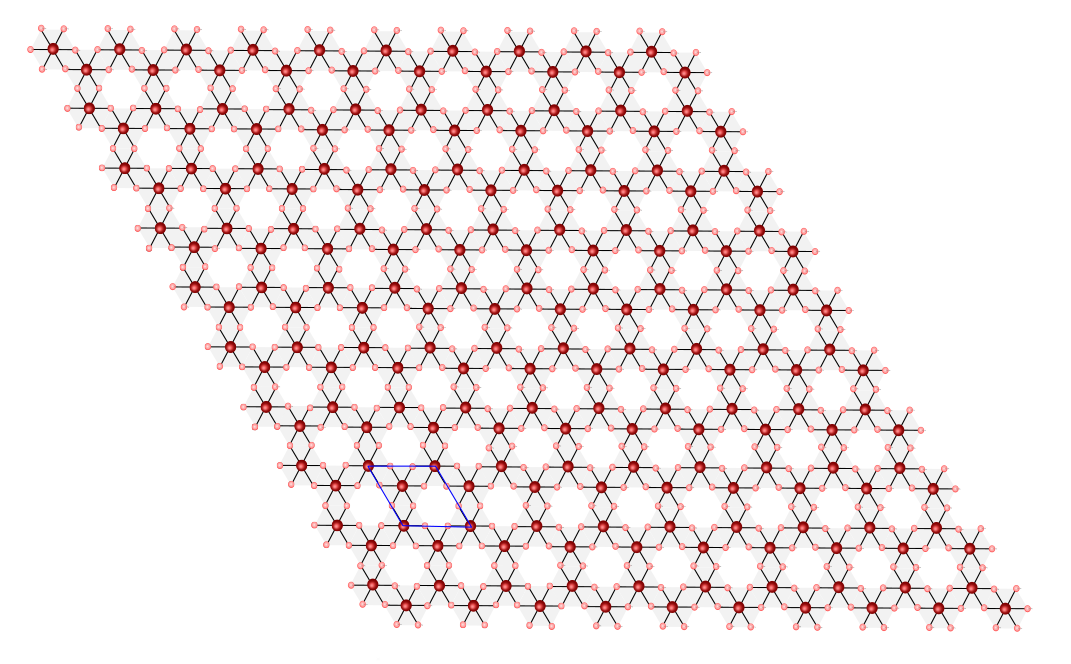}
    \caption{CrI$_3$ structure. The atoms of chromium and iodine are represented by red and pink spheres, respectively. The unit cell of CrI$_3$ consists of two Cr atoms and six I atoms, with the Cr atoms forming a honeycomb structure.}
    \label{fig:lattice}
\end{figure}
\begin{multline}
E_{\textup{m}}=-J\sum_{<i,j>}\boldsymbol{m}_i\cdot\boldsymbol{m}_j-D\sum_{<i,j>}\boldsymbol{d}_{ij} \cdot \left[\boldsymbol{m}_i\times\boldsymbol{m}_j\right]  \\
- K \sum_i\left(\boldsymbol{m}_i\cdot\boldsymbol{e}_z\right)^2-\mu\sum_i \boldsymbol{B}\cdot \boldsymbol{m}_i.
\label{eq:mag_hamiltonian}
\end{multline}
Here, the first, second, third, and fourth terms describe the Heisenberg exchange, the Dzyaloshinskii–Moriya interaction (DMI), the uniaxial magnetocrystalline anisotropy and the Zeeman interaction, respectively; 
$\boldsymbol{m}_i$ is the unit vector pointing along the $i$th magnetic moment, whose magnitude is $\mu$ for each lattice site; $\boldsymbol{B}$ is the external magnetic field, and $\boldsymbol{d}_{ij} = \boldsymbol{r}_{ij}\times \boldsymbol{e}_z/|\boldsymbol{r}_{ij}|$ is the DMI unit vector with $\boldsymbol{e}_z$ and $\boldsymbol{r}_{ij}$ being the unit vector along the monolayer normal and the vector pointing from site $i$ to site $j$, respectively. The angular brackets indicate summation over unique nearest neighbors only. The effective parameter values are taken from Ref.~\cite{ghosh2020comment}: $J=2.53$~meV, $D=1.2$~meV, $K=0.153$~meV, and $\mu=3.0~\mu_B$. The external magnetic field is in the monolayer plane, with the magnitude ranging from 0 T to 3.5 T. In our calculations, we use the computational domain of $N_{\textup{c}}=30\times30=900$ unit cells equipped with periodic boundary conditions. Note that the number of magnetic moments $N_{\textup{a}}$ explicitly included in the calculations is twice as large as $N_c$. 

The Hamiltonian describing the excitonic subsystem reads: 
\begin{multline}
\label{eqn:ex_ham}
H_{\text{exc}}=\sum_{\textbf{q}n}E_{n\textbf{q}}\hat{X}^{\dagger}_{n\textbf{q}}\hat{X}^{}_{n\textbf{q}}\\ +\boldsymbol{E}_{\pm}\sum_{n}\boldsymbol{D}_{n\textbf{q}=0}\hat{X}^{\dagger}_{0\textbf{q}=0}\hat{X}_{n\textbf{k}=0}+h.c.     
\end{multline} 
Here, $\hat{X}^{\dagger}_{n\textbf{q}}$ ($\hat{X}_{n\textbf{q}}$) are the exciton creation (annihilation) operators characterized by momentum $\boldsymbol{q}$ and state $n$ labeling the peak in the absorption spectrum, $E_{n\textbf{q}}$ is the exciton energy, $\boldsymbol{D}_{n\textbf{q}=0}$ is the dipole moment of the optical transition obtained from the DFT calculations (see Appendix~\ref{sec:app_dft}), here limited to the direct-gap transitions,  $\boldsymbol{E}_{\pm}(t)=\text{Re}\left[(E_0,\mp\textup{i}E_0,0)exp(-\textup{i}\omega t)\right]h(t)$ is the right- or left-circularly polarized electric field characterized by the pulse envelope $h(t)$. The effects of both radiative and non-radiative damping are described phenomenologically, see Appendix~\ref{sec:app_damping} for details. 

The interaction between the excitonic and magnetic subsystems is characterized by the following Hamiltonian:
\begin{multline}
   H_{\text{s}}=-g\mu\! \! \!\sum_{\substack{\textbf{q}\textbf{q}'nn'}}\!\!\!\left(\boldsymbol{m}_{\textbf{q}-\textbf{q}'}-\boldsymbol{m}^{\text{g}}_{\textbf{q}-\textbf{q}'}\right)\boldsymbol{M}^{\textbf{q}\textbf{q}'}_{nn'}\hat{X}^{\dagger}_{n\textbf{q}}\hat{X}^{}_{n'\textbf{q}'},
\end{multline}
where the term inside the parentheses is the Fourier transform of the magnetization relative to the collinear ground state (see Appendix~\ref{sec:fourie}), $\boldsymbol{M}^{\textbf{q}\textbf{q}'}_{nn'}$ is the dipole matrix element describing the interaction with local magnetic moment, $g$ is the on-site spin-exciton coupling constant (see Appendix~\ref{sec:app_dft} for the details). The evaluation of the coupling constant $g$ can not be performed using standard DFT approaches and requires a separate consideration which goes beyond the scope of the present work. Here, $g$ is taken as a phenomenological parameter with the value $g=2.3$ meV, which is slightly less then the exchange interaction parameter in the magnetic lattice Hamiltonian. 

Note that exciton-exciton interaction is neglected and only single exciton wavefunction $\Psi_{\text{exc}}$ is considered explicitly. Therefore:  
\begin{align}\label{eqn:lat_ham}
E_{\text{s}}\approx n_{\textup{exc}}\braket{\Psi_{\text{exc}}|H_{\text{s}}|\Psi_{\text{exc}}}=-g\mu \sum\limits_{i}\boldsymbol{\sigma}_i(\boldsymbol{m}_i-\boldsymbol{m}_i^{\textup{g}}).
\end{align}
Here, $n_{\textup{exc}}$ is the scaling factor proportional to the number of the unit cells (in our case $n_{\text{exc}}=450$),  $\boldsymbol{m}^{\textup{g}}\equiv \boldsymbol{e}_z$ is the unit vector along the ground-state magnetization, $\boldsymbol{\sigma}_{i}$ is the exciton spin vector associated with the $i$th unit cell (note that excitons in CrI$_3$ are of the Frenkel type) defined via the following equation:  
\begin{align}
    \boldsymbol{\sigma}_i=n_{\text{exc}}\sum_{\textbf{q}\textbf{q}'}e^{2\pi\textup{i}(\boldsymbol{r}_i(\boldsymbol{q}'-\boldsymbol{q}))_{N_{\textup{c}}}}\sum_{nn'}C_{n}^{\textbf{q}*} \boldsymbol{M}^{\textbf{q}\textbf{q}'}_{nn'}C_{n'}^{\textbf{q}'},\label{eq:sigma}
\end{align}
where $\boldsymbol{r}_i$ is the position of $i$th unit cell and $C_{n}^{\textbf{q}}$ are the expansion coefficients of $\Psi_{\text{exc}}$:
\begin{equation}
\Psi_{\text{exc}}=\sum_{n\textbf{q}}C_{n}^{\textbf{q}} \hat{X}^{\dagger}_{n\textbf{q}} |0\rangle .
\end{equation}
 
\section{Dynamic equations}

The dynamic equations for the observables follow from the model Hamiltonian. The evolution of $\Psi_{\text{exc}}$ is accessed via the equation of motion for its coefficients: 
\begin{equation}
\label{eq:c}
i\hbar\frac{\partial C_{n}^{\textbf{q}}}{\partial t}=\sum_{n'\textbf{q}'} H_{nn'}^{\textbf{q}\textbf{q}'}(t) C_{n'}^{\textbf{q}'}(t),
\end{equation}
where $H_{nn'}^{\textbf{q}\textbf{q}'}$ are the matrix elements of the excitonic Hamiltonian $H=H_{\textup{exc}}+H_{\textup{s}}$.

\begin{figure}[t!]
    \centering
    \includegraphics[width = 1.0\linewidth]{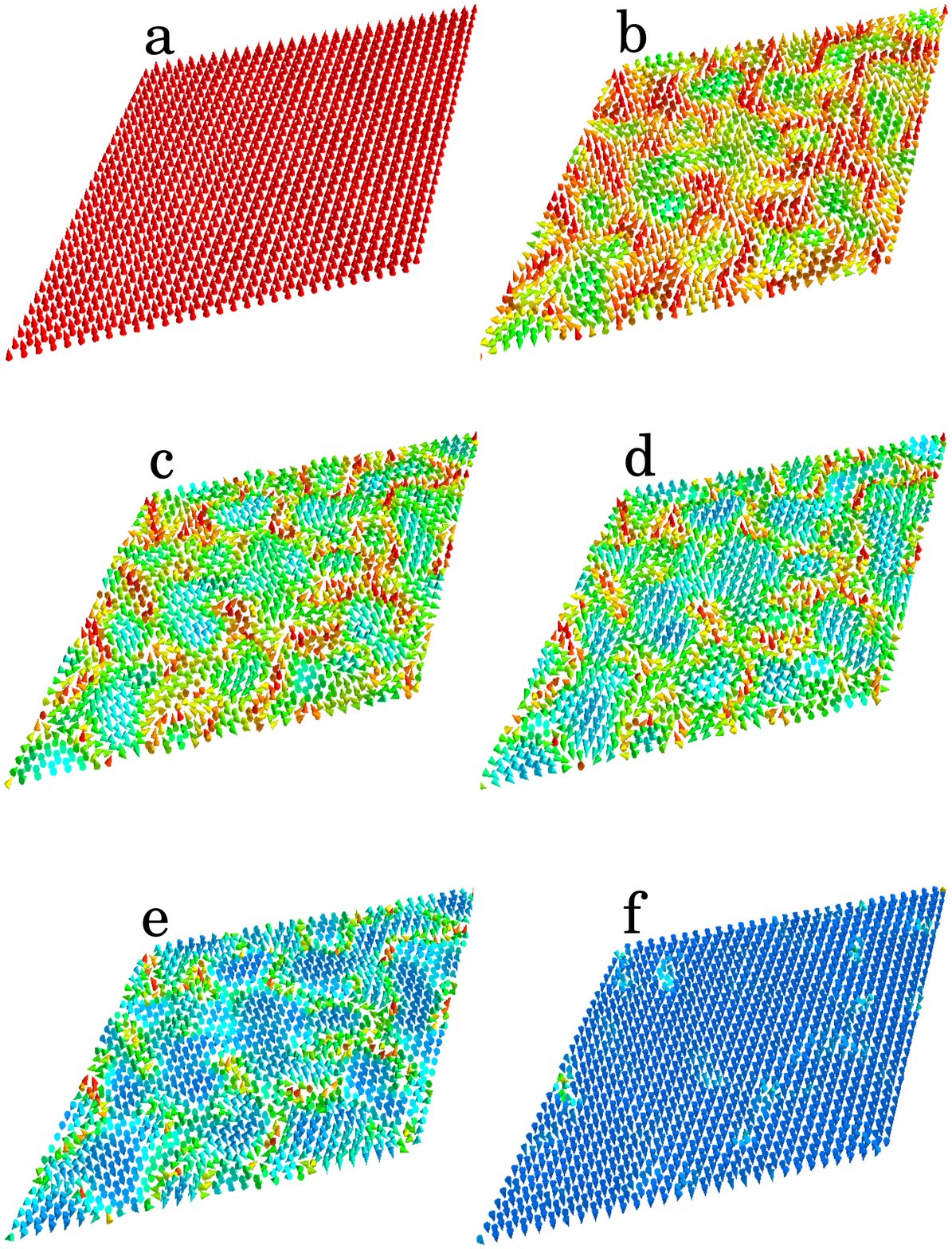}
    \caption{Evolution of the magnetic structure of the system for the pulse fluence $F_1=166$ mJ/cm$^2$. The fluence magnitude is sufficient to induce the magnetization reversal. 
    The snapshots (a)-(f) correspond to $1.6$ ps, $1.8$ ps, $1.84$ ps, $1.86$ ps, $1.9$ ps, and $2.04$ ps, respectively. The color codes the orientation of the magnetic moments of the Cr atoms.} This case corresponds to the solid line in Fig.~\ref{fig:Mz_psi}.
    \label{fig:conf1}
\end{figure}
On the other hand, the dynamics of the lattice magnetization is obtained via the time integration of the Landau-Lifshitz-Gilbert equation (LLGE) for the normalized magnetization vectors at zero temperature:  
\begin{equation}
    \label{eq:llg}
    \dfrac{d\boldsymbol{m}_i}{dt}=-\gamma \boldsymbol{m}_i\times \boldsymbol{B}^{\text{eff}}_{i}+\eta\left(\boldsymbol{m}_i\times\dfrac{d\boldsymbol{m_i}}{dt}\right),
\end{equation}
where $\gamma$ is the gyromagnetic ratio, $\eta$ is the dimensionless damping parameter (in the present study we take $\eta=0.2$), and $\boldsymbol{B}^{\text{eff}}_{i}$ is the effective magnetic field defined as:
\begin{equation}
\label{eq:beff}
    \boldsymbol{B}^{\text{eff}}_{i}=-\frac{1}{\mu}\frac{\partial E}{\partial\boldsymbol{m}_i}. 
\end{equation}
Note that only $E_{\text{m}}+E_{\text{s}}$ depends explicitly on $\boldsymbol{m}_i$. The effect of excitons on the magnetization dynamics is treated within the mean-field approach, where the magnetization-exciton interaction energy $E_{\text{s}}$ depends on the exciton spin vector configuration $\boldsymbol{\sigma}$ [see Eq.~(\ref{eqn:lat_ham})], which needs to be updated every time step using Eqs.~(\ref{eq:c}) and (\ref{eq:sigma}). 
We use the semi-implicit solver by Mentink \textit{et al.} \cite{mentink2010stable} for the time integration of the LLGE. 

\section{Results and discussion}\label{sec:application}

\begin{figure}[b!]
    \centering
    \includegraphics[width = 1.0\linewidth]{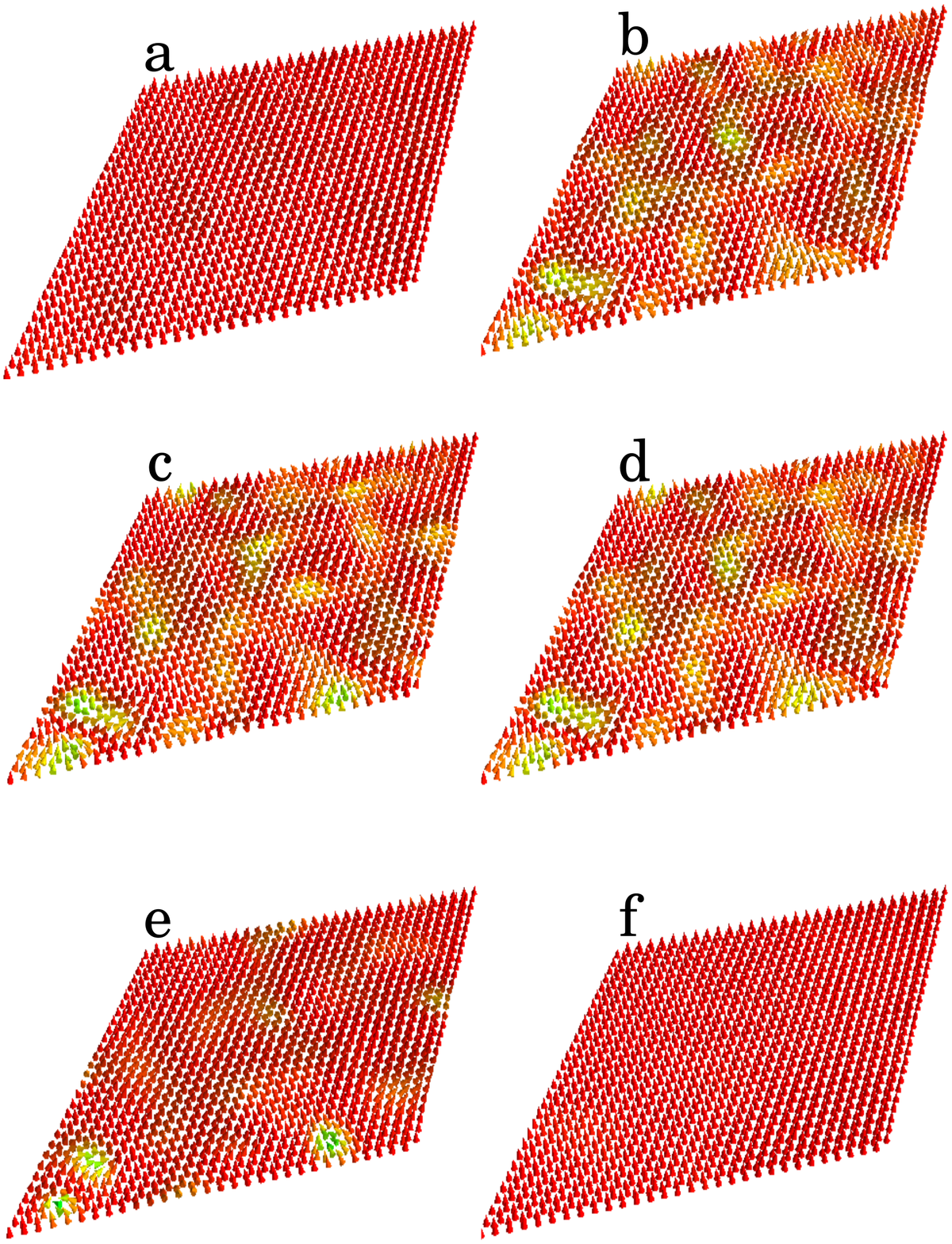}
    \caption{Evolution of the magnetic structure of the system for the pulse fluence $F_2=33$ mJ/cm$^2$. The fluence magnitude is insufficient to induce the magnetization reversal. The snapshots (a)-(f) correspond to $2.2$ ps, $2.76$ ps, $2.98$ ps, $3.2$ ps, $4.4$ ps, and $8.0$ ps, respectively. The color codes the orientation of the magnetic moments of the Cr atoms. This case corresponds to the dashed line in Fig.~\ref{fig:Mz_psi}.}
    \label{unswitch}
\end{figure}

We study the dynamics of the system induced by spatially homogeneous circular polarized laser pulse at normal incidence. The choice of the pulse envelope $h(t)$ is inspired by real setups \cite{Dabrowski2022,zhang2022all}: 
\begin{align}\label{eqn:time_envelope}
h(t) =A \theta(t_f - |2t -t_f|)\exp{\left[-B\left(\dfrac{t-t_f/2}{t_f}\right)^2\right]}.
\end{align}
Here $t_\text{f}$ is the duration of the pulse, and $A=1.94$ and $B=32.2$ are dimensionless parameters defining the total pulse fluence $F$:
\begin{align}\label{eqn:integrated_int}
  F&=\dfrac{E_0^2 c \varepsilon_0}{2} \int\limits_{0}^{t_f}h(t)^2dt=C I_0 t_f ,
\end{align}
where $I_0=E_0^2 c \varepsilon_0$ is the intensity of the pulse and $C=A^2\text{erf}(\sqrt{B/2})\sqrt{\pi/(2B)}$ is the dimensionless constant. In our calculations, the intensity never exceeds the value of 0.1 TW/cm$^2$, which is consistent with experiments~\cite{Dabrowski2022,zhang2022all}. We take the duration of the pulse $t_f=4$ ps and its central frequency $\omega=1.94$ eV, which is below the direct bandgap. Note that the chosen value of $\omega$ is in the vicinity of one of the $\sigma^-$ peaks in the absorption spectrum.

\begin{figure}[t!]
    \centering
    \includegraphics[width = 1.00\linewidth]{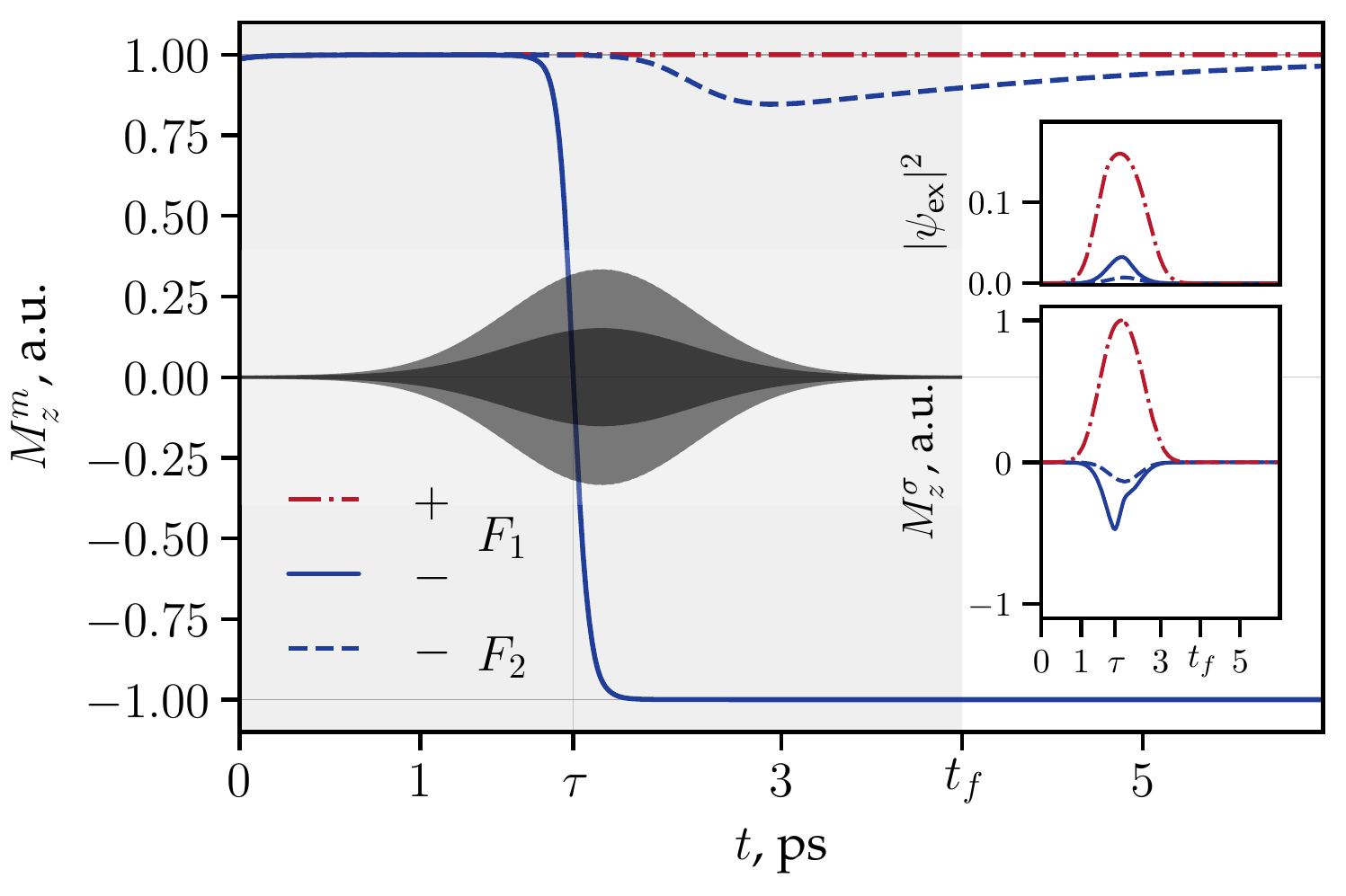}
    \caption{Time-dependence of the cumulative out-of-plane magnetization $M_z^m$ for different light polarizations. Shaded area indicates the laser pulse duration. Upper right inset shows the time dependence of  $|\psi_{ex}|^2 = |\Psi_\text{exc}|^2-|\psi_0|^2$ describing the evolution of excited exciton states. Lower right inset shows the time dependence of the exciton magnetization. Dark grey areas show the envelope functions for the optical pump pulse with fluence $F_1=166$ mJ/cm$^2$ (lighter grey, magnetization reversal occurs) and  $F_2=33.2$ mJ/cm$^2$ (darker grey, magnetization reversal does not occur). The switching time $\tau$ is defined by the instant when $M_z^m$ changes sign. Here $\tau=1.85$ ps. The signs $+$ and $-$ correspond to right-hand and left-hand circularly polarized light, respectively.
    }
    \label{fig:Mz_psi}
\end{figure}

The initial magnetization direction is prepared by aligning all magnetic vectors along $\boldsymbol{e}_z$ and perturbing them by small random noise to break the symmetry. The simulation of the spatio-temporal magnetization pattern provides information about the evolution of the cumulative out-of-plane magnetization due to the spin lattice and excitons: 
\begin{align}
    M_z^m(t)=\dfrac{1}{N_{\textup{a}}}\sum\limits_{i=1}^{N_{\textup{a}}}m_i^z(t), \quad M_z^\sigma(t)=\dfrac{1}{N_{\textup{c}}}\sum\limits_{i=1}^{N_{\textup{c}}}\sigma_i^z(t),
\end{align}
The dynamics of the magnetization switching is shown in Fig.\ref{fig:conf1}.  The arrival of the pulse with $\sigma^-$ polarization leads to the appearance of the randomly distributed domains with inverted magnetization.  
This process is accompanied by the formation of magnetic vortices favored by the DMI. If the fluence exceeds some critical value $F_c$, the size of the domains increases with time, making the system eventually reach the state with spatially homogeneous inverted magnetization (see Fig.~\ref{fig:conf1}). On the other hand, if the fluence is below the critical value, the system relaxes back to the homogeneous state without magnetization inversion (see Fig.~\ref{unswitch})).
\begin{figure}[t!]
    \centering
    \includegraphics[width = 1.0\linewidth]{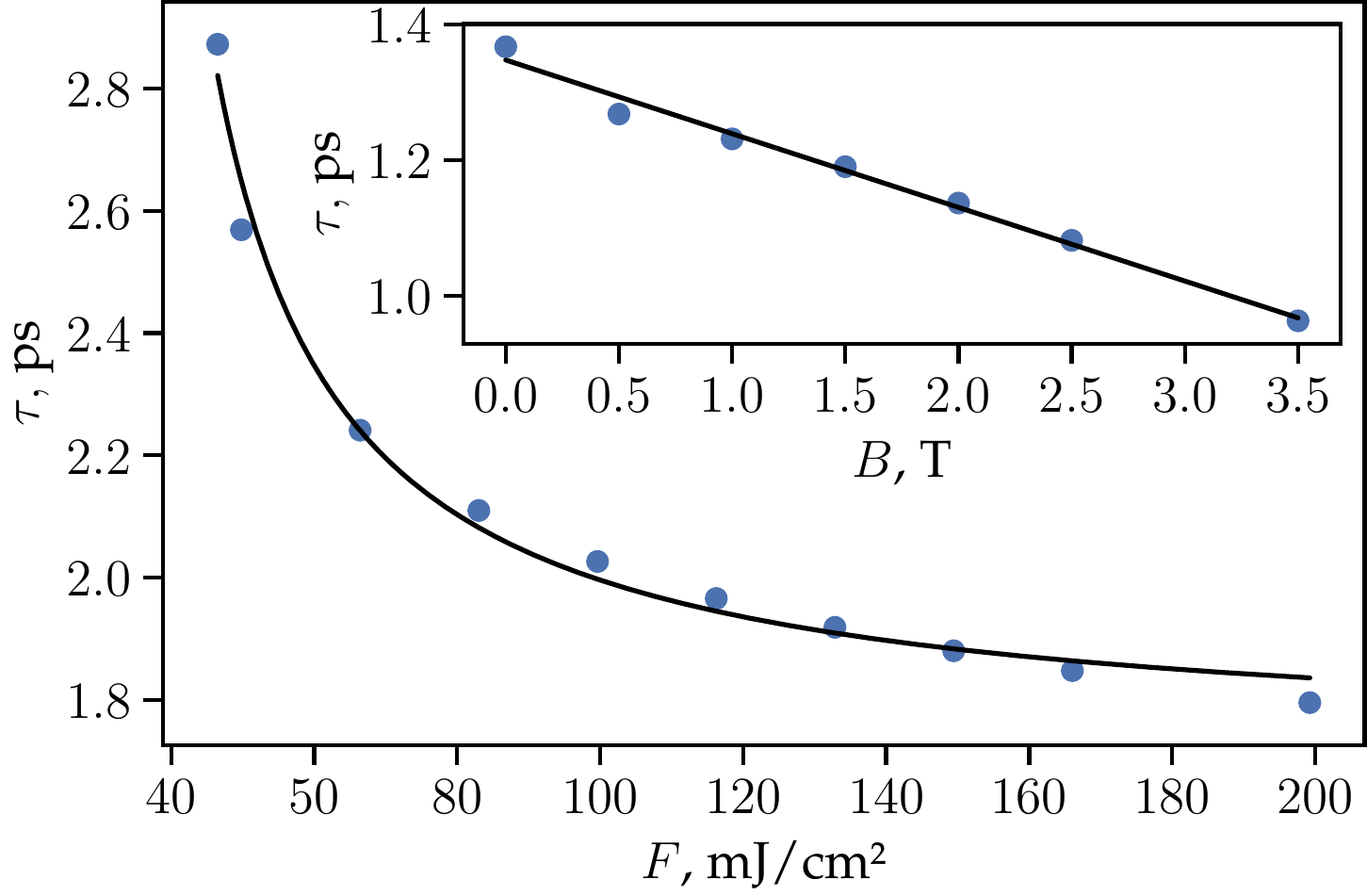}
    \caption{Magnetization switching time $\tau$ as a function of pump fluence $F$. %The influence of pump fluence on transition time. 
    Magnetization switching is achieved if the fluence of the pulse exceeds a critical value $F_c\approx 46$ mJ/cm$^{2}$, at which the switching time diverges. The inset shows the switching time as a function of the lateral external magnetic field for the fluence of $133$ mJ/cm$^2$.}
    \label{fig:fltime}
\end{figure}

The dynamics of the cumulative magnetization is shown in Fig.~\ref{fig:Mz_psi}. The transition between up and down polarized states induced by the pulse is quite abrupt, which makes it possible to introduce the characteristic switching time $\tau$ -- the principal parameter %that 
characterizing the magnetization reversal. As expected, $\tau$ decreases with increasing $F$ (see Fig.~\ref{fig:fltime}). The switching time becomes infinite when the fluence reaches the critical value $F_c\approx 46$ mJ/cm$^{2}$, below which no magnetization reversal is possible. This behavior agrees well with our previous study based on the phenomenological model of resonant magnetization switching~\cite{PhysRevB.104.L020412}. Interestingly, the switching time is influenced by lateral magnetic field (see inset in Fig.~\ref{fig:fltime}).

The change of the circular polarization of the incident beam modifies drastically the magnetization dynamics: the switching does not occur for the initial magnetization pointing up, but can happen for the initial magnetization pointing down.

\section{Conclusion}\label{sec:conclusion}

In conclusion, we developed a microscopic theory of all-optical resonant polarization-sensitive magnetization switching in monolayers of CrI$_3$. The effect is due to the combination of the peculiar optical selection rules for excitons in this material and efficient coupling of excitons to the magnetic lattice. The spatio-temporal distribution of the magnetization under circular polarized pulses was investigated, and the dependence of the parameters characterizing the switching on the properties of the optical pulse was determined.  

\section*{Acknowledgements}
A.K, M.K, and I.A.S. acknowledge the support from the Icelandic Research Fund (Grant No. 163082-051). P.F.B. acknowledges the support from the Icelandic Research Fund (Grant No. 217750), the University of Iceland Research Fund (Grant No. 15673), and the Swedish Research Council (Grant No. 2020-05110). A.K. thanks the University of Iceland for hospitality during the work on the current project. I.V.I and I.A.S acknowledge support  from the joint RFBR-DFG project No. 21-52-12038. The work of A.I.C. and I.V.I. was supported by Rosatom in the framework of the Roadmap for Quantum computing (Contract No. 868-1.3-15/15-2021 dated October 5).  Y.\,V.\,Z. is grateful to the Deutsche Forschungsgemeinschaft (DFG, German Research Foundation) SPP 2244 (Project-ID 443416183) for the financial support.

\appendix

\section{Matrix elements of the exciton Hamiltonian}\label{sec:app_dft}

The matrix elements of the exciton Hamiltonian $H_\text{exc}+H_\text{s}$ are computed via the resolution of the Bethe-Salpeter equation (BSE) parameterized using the first-principles calculations. The exciton wave functions and associated energies are obtained by diagonalizing the BSE Hamiltonian~\cite{RevModPhys.74.601, PhysRevB.62.4927, PhysRevLett.81.2312} as follows:

\begin{align}
    \sum_{c'v'\textbf{k}'}H^{BSE}_{cv\textbf{k}c'v'\textbf{k}'}(\textbf{q})A^{n
    \textbf{q}
    }_{c'v'\textbf{k}'}=E^{n\textbf{q}} A^{n\textbf{q}}_{cv\textbf{q}},
\end{align}
Here, $H^{BSE}_{cv\textbf{k}c'v'\textbf{k}'}(\textbf{q})$ are the matrix elements of the BSE Hamiltonian for excitons possessing momentum $\textbf{q}$, $A^{n\textbf{q}}_{cv\textbf{q}}$ and $E^{n\textbf{q}}$ are the $n$th exciton wave function and energy, respectively. The indices $c$ ($c'$), $v$ ($v'$), and $\textbf{k}$ ($\textbf{k}'$) denote the conduction band, valence band, and single-particle momentum, respectively. The Hamiltonian $H^{BSE}$ is calculated using the Tamm-Dancoff approximation~\cite{RevModPhys.74.601}, which is particularly well-suited for wide-gap semiconductors. This approximation disregards the coupling between resonance and anti-resonance poles while preserving the Hermitian character of the Hamiltonian.

The calculation of the dipole matrix elements is carried out utilizing the following equation:
\begin{align}
    \boldsymbol{D}_{n\textbf{q}=0}=\langle 0 |\boldsymbol{r}| n\textbf{q}=0\rangle=\sum_{cv\textbf{k}}A^{n\textbf{q}=0}_{cv\textbf{k}}\langle v\textbf{k} |\boldsymbol{r}| c\textbf{k} \rangle,
\end{align}
where $\langle v\textbf{k} |\boldsymbol{r}| c\textbf{k} \rangle$  is the single-particle dipole matrix element corresponding to the optical transition from the conduction band $c$ to the valence band $v$ with momentum $\textbf{k}$.

%One of t
The key ingredients of the exciton-skyrmion Hamiltonian interaction are the matrix elements of the exciton magnetic moment $\boldsymbol{M}^{\textbf{q}\textbf{q}'}_{nn'}$.  These matrix elements are assembled from the matrix elements of the spin magnetic moment $\boldsymbol{S}^{\textbf{q}\textbf{q}'}_{nn'}$ and the orbital magnetic moment $\boldsymbol{L}^{\textbf{q}\textbf{q}'}_{nn'}$. The matrix elements of the spin magnetic moment are calculated from the single-particle spin moment:
\begin{align}
\boldsymbol{S}^{\textbf{q}\textbf{q}'}_{nn'}=&\sum_{cv\textbf{k},c'v'\textbf{k}'}\left(A^{n\textbf{q}}_{cv\textbf{k}}\right)^{*}A^{n'\textbf{q}'}_{c'v'\textbf{k}'}\nonumber\\
\qquad\qquad &\times\left[\langle c\textbf{k+q} |\boldsymbol{S}| c'\textbf{k}'+\textbf{q}'\rangle-\langle v'\textbf{k}' |\boldsymbol{S}|  v\textbf{k}\rangle\right],
\end{align}
where $\langle c\textbf{k} |\boldsymbol{S}| c'\textbf{k}' \rangle$ is the single-particle spin matrix element. While the single-particle spin operators are simply defined using the Pauli matrices in the spin subspace, the calculation of the orbital moment requires a more careful consideration. In our study, we employ a local projector augmented wave (PAW) technique to evaluate the orbital single-particle magnetic moment, as it offers a more straightforward approach compared to the previously reported perturbation theory-based methods \cite{PhysRevLett.124.226402, PhysRevB.101.235408, PhysRevB.85.014435, PhysRevLett.95.137205, PhysRevB.74.024408}. All-electron orbitals inside the PAW sphere can be expanded as~\cite{PhysRevB.71.035109, PhysRevB.50.17953}
\begin{align}
    |c\textbf{k}\rangle=\sum_i\langle p_{i}^{a}| \varphi_{c\textbf{k}}\rangle|\phi_{i}^{a}\rangle,
\end{align}
where $p_{i}^{a}$ is a projector of the smooth pseudowave function $\varphi_{c\textbf{k}}$ for band $c$ with momentum $\textbf{k}$ on the $i{\text{th}}$ all-electron partial wave for %$a_{\text{th}}$ 
atom $a$, $ \phi_{i}^{a}$.  We calculate the one-particle matrix elements of the orbital momentum as follows~\cite{PhysRevB.94.235106}:
\begin{align}
    \langle c\textbf{k} |\boldsymbol{L}| c'\textbf{k}' \rangle = \sum_{ai_1i_2}\langle \varphi_{c\textbf{k}} |p_{i_1}^{a}\rangle\langle\phi_{i1}^{a}|\boldsymbol{L}^{a}|\phi_{i2}^{a}\rangle\langle p_{i_2}^{a}|\varphi_{c'\textbf{k}'}\rangle,
\end{align}
where $\boldsymbol{L}^{a}$ is the orbital momentum operator for atom $a$. Since partial waves $\phi_{i}^{a}$ are usually defined in a spherical basis, the matrix elements $\langle\phi_{i1}^{a}|\boldsymbol{L}^{a}|\phi_{i2}^{a}\rangle$ can be calculated analytically. Having single-particle matrix elements of the orbital magnetic moment, we compute the exciton matrix elements as follows:
\begin{align}
\boldsymbol{L}^{\textbf{q}\textbf{q}'}_{nn'}&=\sum_{cv\textbf{k},c'v'\textbf{k}'}\left(A^{n\textbf{q}}_{cv\textbf{k}}\right)^{*}A^{n'\textbf{q}'}_{c'v'\textbf{k}'}\nonumber\\
&\quad\times\left[\langle c\textbf{k+q} |\boldsymbol{L}| c'\textbf{k}'+\textbf{q}' \rangle-\langle v'\textbf{k}' |\boldsymbol{L}|  v\textbf{k}\rangle\right].
\end{align}
The total exciton magnetic moment is obtained by computing a sum of the spin and orbital components:
\begin{align}
    \boldsymbol{M}^{\textbf{q}\textbf{q}'}_{nn'}=\boldsymbol{L}^{\textbf{q}\textbf{q}'}_{nn'}+\boldsymbol{S}^{\textbf{q}\textbf{q}'}_{nn'}.
\end{align}

\section{Introduction of damping}\label{sec:app_damping}

The effects of damping are modeled by introducing finite inverse relaxation time $\delta$, which results in the decay of excited exitonic states $\sim\exp(-\Delta t \delta)$ with time $\Delta t$. For simplicity, the damping parameter is assumed to be the same for all excited excitonic states. Therefore, at each time step of the simulation, the vector of state $\Psi_{\text{exc}}=(\psi_0,\psi_1,\dots,\psi_n)$ with $\psi_0$ being the component responsible for vacuum, is substituted by the modified vector of state $\Psi_{\text{exc}}\leftarrow \tilde{\Psi}_{\text{exc}}=(\tilde{\psi}_{0},\tilde{\psi}_1,\dots,\tilde{\psi}_n)$, whose componens are defined via the following operations:
\begin{align}
    & \tilde{\psi}_{i}=\psi_i\sqrt{1-\Delta t \delta},\text{} i>0\\
    &\Delta = 1 - \left(\sum\limits_{i=1}^n|\tilde{\psi}_{i}|^2 +|\psi_{0}|^2\right),\\
    & \tilde{\psi}_0=\psi_0\dfrac{\sqrt{\Delta + |\psi_0|^2}}{|\psi_0|}.
\end{align}
Note that $|\tilde{\Psi}_{\text{exc}}|=|\Psi_\text{exc}|=1$. In our study, we take $\delta=80$ meV.

\section{Fourier transforms}\label{sec:fourie}

The definition of the direct and inverse Fourier transforms used in this study requires a special discussion since %a are given in this Section.  
%It is slightly different from the standard notation, the thing is that 
the gratings in the real space(r) and momentum space(q) have different discreteness. %The function $f(q_j)$ can be selected as an optional one. 
For an arbitrary function $f$, the Fourier transforms are defined via the following equations:
\begin{align}
f(\boldsymbol{r}_i) =& \sum\limits_{j}^{N_q} \exp\left[2\pi\text{i}(\boldsymbol{r}_i\boldsymbol{q}_j)_{N_c} \right] f(\boldsymbol{q}_j),\\
f(\boldsymbol{q}_j) =& \dfrac{1}{N_c}\sum\limits_{i}^{N_{\textup{c}}} \exp\left[-2\pi\text{i}(\boldsymbol{r}_i\boldsymbol{q}_j)_{N_c}\right]f(\boldsymbol{r}_i),
\end{align}
where $N_{\textup{c}}=30\times30=900$ is the number of unit cells and $N_q=6\times6=36$ is the number of points in the $q$-space. We define the dot product $(\dots)_{N_c}$ as:
\begin{align}
(\boldsymbol{r}\boldsymbol{q})_{N_c}=\frac{r^x _i q^x _j}{N^x_c}+\frac{r^y _i q^y _j}{N^y_c}+\frac{r^z_i q^z_j}{N^z_c}.
\end{align}
We use $N^x_{\textup{c}}=N^y_{\textup{c}}=30$ and $N^z_{\textup{c}}=1$. %Let us give a little comment on the calculation of $\boldsymbol{m}\left(\textbf{q}-\textbf{q}'\right)$. 
Since there are two magnetic atoms per unit cell, we adhere to the following formula to compute the Fourier transform of the magnetization:
\begin{align}
    &\boldsymbol{m}_{\textbf{q}-\textbf{q}'}=\dfrac{1}{N_c}\sum\limits_{i}^{N_{\textup{c}}} \exp\left[-2\pi\text{i}(\boldsymbol{r}_i\left(\textbf{q}-\textbf{q}'\right))_{N_c}\right]\\ \nonumber
    &\qquad\qquad\qquad\qquad\qquad\qquad \times (\boldsymbol{m}_{2i}+\boldsymbol{m}_{2i-1}),
\end{align}
where we use the total magnetic moment of the unit cell.

%\bibliography{lit.bib}

\begin{thebibliography}{38}%
\makeatletter
\providecommand \@ifxundefined [1]{%
 \@ifx{#1\undefined}
}%
\providecommand \@ifnum [1]{%
 \ifnum #1\expandafter \@firstoftwo
 \else \expandafter \@secondoftwo
 \fi
}%
\providecommand \@ifx [1]{%
 \ifx #1\expandafter \@firstoftwo
 \else \expandafter \@secondoftwo
 \fi
}%
\providecommand \natexlab [1]{#1}%
\providecommand \enquote  [1]{``#1''}%
\providecommand \bibnamefont  [1]{#1}%
\providecommand \bibfnamefont [1]{#1}%
\providecommand \citenamefont [1]{#1}%
\providecommand \href@noop [0]{\@secondoftwo}%
\providecommand \href [0]{\begingroup \@sanitize@url \@href}%
\providecommand \@href[1]{\@@startlink{#1}\@@href}%
\providecommand \@@href[1]{\endgroup#1\@@endlink}%
\providecommand \@sanitize@url [0]{\catcode `\\12\catcode `\$12\catcode
  `\&12\catcode `\#12\catcode `\^12\catcode `\_12\catcode `\%12\relax}%
\providecommand \@@startlink[1]{}%
\providecommand \@@endlink[0]{}%
\providecommand \url  [0]{\begingroup\@sanitize@url \@url }%
\providecommand \@url [1]{\endgroup\@href {#1}{\urlprefix }}%
\providecommand \urlprefix  [0]{URL }%
\providecommand \Eprint [0]{\href }%
\providecommand \doibase [0]{http://dx.doi.org/}%
\providecommand \selectlanguage [0]{\@gobble}%
\providecommand \bibinfo  [0]{\@secondoftwo}%
\providecommand \bibfield  [0]{\@secondoftwo}%
\providecommand \translation [1]{[#1]}%
\providecommand \BibitemOpen [0]{}%
\providecommand \bibitemStop [0]{}%
\providecommand \bibitemNoStop [0]{.\EOS\space}%
\providecommand \EOS [0]{\spacefactor3000\relax}%
\providecommand \BibitemShut  [1]{\csname bibitem#1\endcsname}%
\let\auto@bib@innerbib\@empty
%</preamble>
\bibitem [{\citenamefont {Gorchon}\ \emph {et~al.}(2016)\citenamefont
  {Gorchon}, \citenamefont {Wilson}, \citenamefont {Yang}, \citenamefont
  {Pattabi}, \citenamefont {Chen}, \citenamefont {He}, \citenamefont {Wang},
  \citenamefont {Li},\ and\ \citenamefont {Bokor}}]{PhysRevB.94.184406}%
  \BibitemOpen
  \bibfield  {author} {\bibinfo {author} {\bibfnamefont {J.}~\bibnamefont
  {Gorchon}}, \bibinfo {author} {\bibfnamefont {R.~B.}\ \bibnamefont {Wilson}},
  \bibinfo {author} {\bibfnamefont {Y.}~\bibnamefont {Yang}}, \bibinfo {author}
  {\bibfnamefont {A.}~\bibnamefont {Pattabi}}, \bibinfo {author} {\bibfnamefont
  {J.~Y.}\ \bibnamefont {Chen}}, \bibinfo {author} {\bibfnamefont
  {L.}~\bibnamefont {He}}, \bibinfo {author} {\bibfnamefont {J.~P.}\
  \bibnamefont {Wang}}, \bibinfo {author} {\bibfnamefont {M.}~\bibnamefont
  {Li}}, \ and\ \bibinfo {author} {\bibfnamefont {J.}~\bibnamefont {Bokor}},\
  }\bibfield  {title} {\enquote {\bibinfo {title} {Role of electron and phonon
  temperatures in the helicity-independent all-optical switching of gdfeco},}\
  }\href {\doibase 10.1103/PhysRevB.94.184406} {\bibfield  {journal} {\bibinfo
  {journal} {Phys. Rev. B}\ }\textbf {\bibinfo {volume} {94}},\ \bibinfo
  {pages} {184406} (\bibinfo {year} {2016})}\BibitemShut {NoStop}%
\bibitem [{\citenamefont {Moreno}\ \emph {et~al.}(2017)\citenamefont {Moreno},
  \citenamefont {Ostler}, \citenamefont {Chantrell},\ and\ \citenamefont
  {Chubykalo-Fesenko}}]{PhysRevB.96.014409}%
  \BibitemOpen
  \bibfield  {author} {\bibinfo {author} {\bibfnamefont {R.}~\bibnamefont
  {Moreno}}, \bibinfo {author} {\bibfnamefont {T.~A.}\ \bibnamefont {Ostler}},
  \bibinfo {author} {\bibfnamefont {R.~W.}\ \bibnamefont {Chantrell}}, \ and\
  \bibinfo {author} {\bibfnamefont {O.}~\bibnamefont {Chubykalo-Fesenko}},\
  }\bibfield  {title} {\enquote {\bibinfo {title} {Conditions for thermally
  induced all-optical switching in ferrimagnetic alloys: Modeling of tbco},}\
  }\href {\doibase 10.1103/PhysRevB.96.014409} {\bibfield  {journal} {\bibinfo
  {journal} {Phys. Rev. B}\ }\textbf {\bibinfo {volume} {96}},\ \bibinfo
  {pages} {014409} (\bibinfo {year} {2017})}\BibitemShut {NoStop}%
\bibitem [{\citenamefont {Ignatyeva}\ \emph {et~al.}(2019)\citenamefont
  {Ignatyeva}, \citenamefont {Davies}, \citenamefont {Sylgacheva},
  \citenamefont {Tsukamoto}, \citenamefont {Yoshikawa}, \citenamefont
  {Kapralov}, \citenamefont {Kirilyuk}, \citenamefont {Belotelov},\ and\
  \citenamefont {Kimel}}]{Ignatyeva2019}%
  \BibitemOpen
  \bibfield  {author} {\bibinfo {author} {\bibfnamefont {D.~O.}\ \bibnamefont
  {Ignatyeva}}, \bibinfo {author} {\bibfnamefont {C.~S.}\ \bibnamefont
  {Davies}}, \bibinfo {author} {\bibfnamefont {D.~A.}\ \bibnamefont
  {Sylgacheva}}, \bibinfo {author} {\bibfnamefont {A.}~\bibnamefont
  {Tsukamoto}}, \bibinfo {author} {\bibfnamefont {H.}~\bibnamefont
  {Yoshikawa}}, \bibinfo {author} {\bibfnamefont {P.~O.}\ \bibnamefont
  {Kapralov}}, \bibinfo {author} {\bibfnamefont {A.}~\bibnamefont {Kirilyuk}},
  \bibinfo {author} {\bibfnamefont {V.~I.}\ \bibnamefont {Belotelov}}, \ and\
  \bibinfo {author} {\bibfnamefont {A.~V.}\ \bibnamefont {Kimel}},\ }\bibfield
  {title} {\enquote {\bibinfo {title} {Plasmonic layer-selective all-optical
  switching of magnetization with nanometer resolution},}\ }\href@noop {}
  {\bibfield  {journal} {\bibinfo  {journal} {Nature Comm.}\ }\textbf {\bibinfo
  {volume} {10}},\ \bibinfo {pages} {4786} (\bibinfo {year}
  {2019})}\BibitemShut {NoStop}%
\bibitem [{\citenamefont {Aviles-Flix}\ \emph {et~al.}(2020)\citenamefont
  {Aviles-Flix}, \citenamefont {Olivier}, \citenamefont {Li}, \citenamefont
  {Davies}, \citenamefont {Alvaro-Gomez}, \citenamefont {Rubio-Roy},
  \citenamefont {Auffret}, \citenamefont {Kirilyuk}, \citenamefont {Kimel},
  \citenamefont {Rasing}, \citenamefont {Buda-Prejbeanu}, \citenamefont
  {Sousa}, \citenamefont {Dieny},\ and\ \citenamefont
  {Prejbeanu}}]{Alives2020}%
  \BibitemOpen
  \bibfield  {author} {\bibinfo {author} {\bibfnamefont {L.}~\bibnamefont
  {Aviles-Flix}}, \bibinfo {author} {\bibfnamefont {A.}~\bibnamefont
  {Olivier}}, \bibinfo {author} {\bibfnamefont {G.}~\bibnamefont {Li}},
  \bibinfo {author} {\bibfnamefont {C.~S.}\ \bibnamefont {Davies}}, \bibinfo
  {author} {\bibfnamefont {L.}~\bibnamefont {Alvaro-Gomez}}, \bibinfo {author}
  {\bibfnamefont {M.}~\bibnamefont {Rubio-Roy}}, \bibinfo {author}
  {\bibfnamefont {S.}~\bibnamefont {Auffret}}, \bibinfo {author} {\bibfnamefont
  {A.}~\bibnamefont {Kirilyuk}}, \bibinfo {author} {\bibfnamefont {A.~V.}\
  \bibnamefont {Kimel}}, \bibinfo {author} {\bibfnamefont {Th.}\ \bibnamefont
  {Rasing}}, \bibinfo {author} {\bibfnamefont {L.~D.}\ \bibnamefont
  {Buda-Prejbeanu}}, \bibinfo {author} {\bibfnamefont {R.~C.}\ \bibnamefont
  {Sousa}}, \bibinfo {author} {\bibfnamefont {B.}~\bibnamefont {Dieny}}, \ and\
  \bibinfo {author} {\bibfnamefont {I.~L.}\ \bibnamefont {Prejbeanu}},\
  }\bibfield  {title} {\enquote {\bibinfo {title} {Single-shot all-optical
  switching of magnetization in tb/co multilayer-based electrodes},}\
  }\href@noop {} {\bibfield  {journal} {\bibinfo  {journal} {Sci. Rep.}\
  }\textbf {\bibinfo {volume} {10}},\ \bibinfo {pages} {5211} (\bibinfo {year}
  {2020})}\BibitemShut {NoStop}%
\bibitem [{\citenamefont {Stanciu}\ \emph {et~al.}(2007)\citenamefont
  {Stanciu}, \citenamefont {Hansteen}, \citenamefont {Kimel}, \citenamefont
  {Kirilyuk}, \citenamefont {Tsukamoto}, \citenamefont {Itoh},\ and\
  \citenamefont {Rasing}}]{Stanciu2020}%
  \BibitemOpen
  \bibfield  {author} {\bibinfo {author} {\bibfnamefont {C.~D.}\ \bibnamefont
  {Stanciu}}, \bibinfo {author} {\bibfnamefont {F.}~\bibnamefont {Hansteen}},
  \bibinfo {author} {\bibfnamefont {A.~V.}\ \bibnamefont {Kimel}}, \bibinfo
  {author} {\bibfnamefont {A.}~\bibnamefont {Kirilyuk}}, \bibinfo {author}
  {\bibfnamefont {A.}~\bibnamefont {Tsukamoto}}, \bibinfo {author}
  {\bibfnamefont {A.}~\bibnamefont {Itoh}}, \ and\ \bibinfo {author}
  {\bibfnamefont {Th.}\ \bibnamefont {Rasing}},\ }\bibfield  {title} {\enquote
  {\bibinfo {title} {All-optical magnetic recording with circularly polarized
  light},}\ }\href@noop {} {\bibfield  {journal} {\bibinfo  {journal} {Phys.
  Rev. Lett.}\ }\textbf {\bibinfo {volume} {99}},\ \bibinfo {pages} {047601}
  (\bibinfo {year} {2007})}\BibitemShut {NoStop}%
\bibitem [{\citenamefont {Davies}\ \emph {et~al.}(2020)\citenamefont {Davies},
  \citenamefont {Janssen}, \citenamefont {Mentink}, \citenamefont {Tsukamoto},
  \citenamefont {Kimel}, \citenamefont {van~der Meer}, \citenamefont
  {Stupakiewicz},\ and\ \citenamefont {Kirilyuk}}]{Davies2020}%
  \BibitemOpen
  \bibfield  {author} {\bibinfo {author} {\bibfnamefont {C.~S.}\ \bibnamefont
  {Davies}}, \bibinfo {author} {\bibfnamefont {T.}~\bibnamefont {Janssen}},
  \bibinfo {author} {\bibfnamefont {J.~H.}\ \bibnamefont {Mentink}}, \bibinfo
  {author} {\bibfnamefont {A.}~\bibnamefont {Tsukamoto}}, \bibinfo {author}
  {\bibfnamefont {A.~V.}\ \bibnamefont {Kimel}}, \bibinfo {author}
  {\bibfnamefont {A.~F.~G.}\ \bibnamefont {van~der Meer}}, \bibinfo {author}
  {\bibfnamefont {A.}~\bibnamefont {Stupakiewicz}}, \ and\ \bibinfo {author}
  {\bibfnamefont {A.}~\bibnamefont {Kirilyuk}},\ }\bibfield  {title} {\enquote
  {\bibinfo {title} {Pathways for single-shot all-optical switching of
  magnetization in ferrimagnets},}\ }\href@noop {} {\bibfield  {journal}
  {\bibinfo  {journal} {Phys. Rev. Applied}\ }\textbf {\bibinfo {volume}
  {13}},\ \bibinfo {pages} {024064} (\bibinfo {year} {2020})}\BibitemShut
  {NoStop}%
\bibitem [{\citenamefont {Igarashi}\ \emph {et~al.}(2020)\citenamefont
  {Igarashi}, \citenamefont {Remy}, \citenamefont {Iihama}, \citenamefont
  {Malinowski}, \citenamefont {Hehn}, \citenamefont {Gorchon}, \citenamefont
  {Hohlfeld}, \citenamefont {Fukami}, \citenamefont {Ohno},\ and\ \citenamefont
  {Mangin}}]{Igrashi2020}%
  \BibitemOpen
  \bibfield  {author} {\bibinfo {author} {\bibfnamefont {J.}~\bibnamefont
  {Igarashi}}, \bibinfo {author} {\bibfnamefont {Q.}~\bibnamefont {Remy}},
  \bibinfo {author} {\bibfnamefont {S.}~\bibnamefont {Iihama}}, \bibinfo
  {author} {\bibfnamefont {G.}~\bibnamefont {Malinowski}}, \bibinfo {author}
  {\bibfnamefont {M.}~\bibnamefont {Hehn}}, \bibinfo {author} {\bibfnamefont
  {J.}~\bibnamefont {Gorchon}}, \bibinfo {author} {\bibfnamefont
  {J.}~\bibnamefont {Hohlfeld}}, \bibinfo {author} {\bibfnamefont
  {S.}~\bibnamefont {Fukami}}, \bibinfo {author} {\bibfnamefont
  {H.}~\bibnamefont {Ohno}}, \ and\ \bibinfo {author} {\bibfnamefont
  {S.}~\bibnamefont {Mangin}},\ }\bibfield  {title} {\enquote {\bibinfo {title}
  {Engineering single-shot all-optical switching of ferromagnetic materials},}\
  }\href@noop {} {\bibfield  {journal} {\bibinfo  {journal} {Nano Lett.}\
  }\textbf {\bibinfo {volume} {20}},\ \bibinfo {pages} {8654} (\bibinfo {year}
  {2020})}\BibitemShut {NoStop}%
\bibitem [{\citenamefont {Lu}\ \emph {et~al.}(2018)\citenamefont {Lu},
  \citenamefont {Zou}, \citenamefont {Hinzke}, \citenamefont {Liu},
  \citenamefont {Wang}, \citenamefont {Cheng}, \citenamefont {Wu},
  \citenamefont {Ostler}, \citenamefont {Cai}, \citenamefont {Nowak},
  \citenamefont {Chantrell}, \citenamefont {Zhai},\ and\ \citenamefont
  {Xu}}]{Lu2018}%
  \BibitemOpen
  \bibfield  {author} {\bibinfo {author} {\bibfnamefont {X.}~\bibnamefont
  {Lu}}, \bibinfo {author} {\bibfnamefont {X.}~\bibnamefont {Zou}}, \bibinfo
  {author} {\bibfnamefont {D.}~\bibnamefont {Hinzke}}, \bibinfo {author}
  {\bibfnamefont {T.}~\bibnamefont {Liu}}, \bibinfo {author} {\bibfnamefont
  {Y.}~\bibnamefont {Wang}}, \bibinfo {author} {\bibfnamefont {T.}~\bibnamefont
  {Cheng}}, \bibinfo {author} {\bibfnamefont {J.}~\bibnamefont {Wu}}, \bibinfo
  {author} {\bibfnamefont {T.~A.}\ \bibnamefont {Ostler}}, \bibinfo {author}
  {\bibfnamefont {J.}~\bibnamefont {Cai}}, \bibinfo {author} {\bibfnamefont
  {U.}~\bibnamefont {Nowak}}, \bibinfo {author} {\bibfnamefont {R.~W.}\
  \bibnamefont {Chantrell}}, \bibinfo {author} {\bibfnamefont {Y.}~\bibnamefont
  {Zhai}}, \ and\ \bibinfo {author} {\bibfnamefont {Y.}~\bibnamefont {Xu}},\
  }\bibfield  {title} {\enquote {\bibinfo {title} {Roles of heating and
  helicity in ultrafast all-optical magnetization switching in tbfeco},}\
  }\href@noop {} {\bibfield  {journal} {\bibinfo  {journal} {Appl. Phys.
  Lett.}\ }\textbf {\bibinfo {volume} {113}},\ \bibinfo {pages} {032405}
  (\bibinfo {year} {2018})}\BibitemShut {NoStop}%
\bibitem [{\citenamefont {El~Hadri}\ \emph {et~al.}(2016)\citenamefont
  {El~Hadri}, \citenamefont {Pirro}, \citenamefont {Lambert}, \citenamefont
  {Petit-Watelot}, \citenamefont {Quessab}, \citenamefont {Hehn}, \citenamefont
  {Montaigne}, \citenamefont {Malinowski},\ and\ \citenamefont
  {Mangin}}]{PhysRevB.94.064412}%
  \BibitemOpen
  \bibfield  {author} {\bibinfo {author} {\bibfnamefont {M.~S.}\ \bibnamefont
  {El~Hadri}}, \bibinfo {author} {\bibfnamefont {P.}~\bibnamefont {Pirro}},
  \bibinfo {author} {\bibfnamefont {C.-H.}\ \bibnamefont {Lambert}}, \bibinfo
  {author} {\bibfnamefont {S.}~\bibnamefont {Petit-Watelot}}, \bibinfo {author}
  {\bibfnamefont {Y.}~\bibnamefont {Quessab}}, \bibinfo {author} {\bibfnamefont
  {M.}~\bibnamefont {Hehn}}, \bibinfo {author} {\bibfnamefont {F.}~\bibnamefont
  {Montaigne}}, \bibinfo {author} {\bibfnamefont {G.}~\bibnamefont
  {Malinowski}}, \ and\ \bibinfo {author} {\bibfnamefont {S.}~\bibnamefont
  {Mangin}},\ }\bibfield  {title} {\enquote {\bibinfo {title} {Two types of
  all-optical magnetization switching mechanisms using femtosecond laser
  pulses},}\ }\href {\doibase 10.1103/PhysRevB.94.064412} {\bibfield  {journal}
  {\bibinfo  {journal} {Phys. Rev. B}\ }\textbf {\bibinfo {volume} {94}},\
  \bibinfo {pages} {064412} (\bibinfo {year} {2016})}\BibitemShut {NoStop}%
\bibitem [{\citenamefont {van Hees}\ \emph {et~al.}(2020)\citenamefont {van
  Hees}, \citenamefont {van~de Meugheuvel}, \citenamefont {Koopmans},\ and\
  \citenamefont {Lavrijsen}}]{vanHees2020}%
  \BibitemOpen
  \bibfield  {author} {\bibinfo {author} {\bibfnamefont {Y.~L.~W.}\
  \bibnamefont {van Hees}}, \bibinfo {author} {\bibfnamefont {P.}~\bibnamefont
  {van~de Meugheuvel}}, \bibinfo {author} {\bibfnamefont {B.}~\bibnamefont
  {Koopmans}}, \ and\ \bibinfo {author} {\bibfnamefont {R.}~\bibnamefont
  {Lavrijsen}},\ }\bibfield  {title} {\enquote {\bibinfo {title} {Deterministic
  all-optical magnetization writing facilitated by non-local transfer of spin
  angular momentum},}\ }\href@noop {} {\bibfield  {journal} {\bibinfo
  {journal} {Nature Comm.}\ }\textbf {\bibinfo {volume} {11}},\ \bibinfo
  {pages} {3835} (\bibinfo {year} {2020})}\BibitemShut {NoStop}%
\bibitem [{\citenamefont {Zhang}\ \emph {et~al.}(2022)\citenamefont {Zhang},
  \citenamefont {Chung}, \citenamefont {Li}, \citenamefont {Wang},
  \citenamefont {Wang}, \citenamefont {Huey}, \citenamefont {Yang},
  \citenamefont {Goldberger}, \citenamefont {Yao},\ and\ \citenamefont
  {Zhang}}]{zhang2022all}%
  \BibitemOpen
  \bibfield  {author} {\bibinfo {author} {\bibfnamefont {Peiyao}\ \bibnamefont
  {Zhang}}, \bibinfo {author} {\bibfnamefont {Ting-Fung}\ \bibnamefont
  {Chung}}, \bibinfo {author} {\bibfnamefont {Quanwei}\ \bibnamefont {Li}},
  \bibinfo {author} {\bibfnamefont {Siqi}\ \bibnamefont {Wang}}, \bibinfo
  {author} {\bibfnamefont {Qingjun}\ \bibnamefont {Wang}}, \bibinfo {author}
  {\bibfnamefont {Warren L.~B.}\ \bibnamefont {Huey}}, \bibinfo {author}
  {\bibfnamefont {Sui}\ \bibnamefont {Yang}}, \bibinfo {author} {\bibfnamefont
  {Joshua~E.}\ \bibnamefont {Goldberger}}, \bibinfo {author} {\bibfnamefont
  {Jie}\ \bibnamefont {Yao}}, \ and\ \bibinfo {author} {\bibfnamefont {Xiang}\
  \bibnamefont {Zhang}},\ }\bibfield  {title} {\enquote {\bibinfo {title}
  {All-optical switching of magnetization in atomically thin {CrI}3},}\ }\href
  {\doibase 10.1038/s41563-022-01354-7} {\bibfield  {journal} {\bibinfo
  {journal} {Nature Materials}\ }\textbf {\bibinfo {volume} {21}},\ \bibinfo
  {pages} {1373--1378} (\bibinfo {year} {2022})}\BibitemShut {NoStop}%
\bibitem [{\citenamefont {Huang}\ \emph {et~al.}(2018)\citenamefont {Huang},
  \citenamefont {Clark}, \citenamefont {Klein}, \citenamefont {MacNeill},
  \citenamefont {Navarro-Moratalla}, \citenamefont {Seyler}, \citenamefont
  {Wilson}, \citenamefont {McGuire}, \citenamefont {Cobden}, \citenamefont
  {Xiao} \emph {et~al.}}]{huang2018electrical}%
  \BibitemOpen
  \bibfield  {author} {\bibinfo {author} {\bibfnamefont {Bevin}\ \bibnamefont
  {Huang}}, \bibinfo {author} {\bibfnamefont {Genevieve}\ \bibnamefont
  {Clark}}, \bibinfo {author} {\bibfnamefont {Dahlia~R}\ \bibnamefont {Klein}},
  \bibinfo {author} {\bibfnamefont {David}\ \bibnamefont {MacNeill}}, \bibinfo
  {author} {\bibfnamefont {Efr{\'e}n}\ \bibnamefont {Navarro-Moratalla}},
  \bibinfo {author} {\bibfnamefont {Kyle~L}\ \bibnamefont {Seyler}}, \bibinfo
  {author} {\bibfnamefont {Nathan}\ \bibnamefont {Wilson}}, \bibinfo {author}
  {\bibfnamefont {Michael~A}\ \bibnamefont {McGuire}}, \bibinfo {author}
  {\bibfnamefont {David~H}\ \bibnamefont {Cobden}}, \bibinfo {author}
  {\bibfnamefont {Di}~\bibnamefont {Xiao}},  \emph {et~al.},\ }\bibfield
  {title} {\enquote {\bibinfo {title} {Electrical control of 2d magnetism in
  bilayer cri3},}\ }\href@noop {} {\bibfield  {journal} {\bibinfo  {journal}
  {Nature nanotechnology}\ }\textbf {\bibinfo {volume} {13}},\ \bibinfo {pages}
  {544--548} (\bibinfo {year} {2018})}\BibitemShut {NoStop}%
\bibitem [{\citenamefont {McGuire}\ \emph {et~al.}(2015)\citenamefont
  {McGuire}, \citenamefont {Dixit}, \citenamefont {Cooper},\ and\ \citenamefont
  {Sales}}]{mcguire2015coupling}%
  \BibitemOpen
  \bibfield  {author} {\bibinfo {author} {\bibfnamefont {Michael~A}\
  \bibnamefont {McGuire}}, \bibinfo {author} {\bibfnamefont {Hemant}\
  \bibnamefont {Dixit}}, \bibinfo {author} {\bibfnamefont {Valentino~R}\
  \bibnamefont {Cooper}}, \ and\ \bibinfo {author} {\bibfnamefont {Brian~C}\
  \bibnamefont {Sales}},\ }\bibfield  {title} {\enquote {\bibinfo {title}
  {Coupling of crystal structure and magnetism in the layered, ferromagnetic
  insulator cri3},}\ }\href@noop {} {\bibfield  {journal} {\bibinfo  {journal}
  {Chemistry of Materials}\ }\textbf {\bibinfo {volume} {27}},\ \bibinfo
  {pages} {612--620} (\bibinfo {year} {2015})}\BibitemShut {NoStop}%
\bibitem [{\citenamefont {Wu}\ \emph {et~al.}(2019)\citenamefont {Wu},
  \citenamefont {Li}, \citenamefont {Cao},\ and\ \citenamefont
  {Louie}}]{Wu2019}%
  \BibitemOpen
  \bibfield  {author} {\bibinfo {author} {\bibfnamefont {M.}~\bibnamefont
  {Wu}}, \bibinfo {author} {\bibfnamefont {Z.}~\bibnamefont {Li}}, \bibinfo
  {author} {\bibfnamefont {T.}~\bibnamefont {Cao}}, \ and\ \bibinfo {author}
  {\bibfnamefont {S.~G.}\ \bibnamefont {Louie}},\ }\bibfield  {title} {\enquote
  {\bibinfo {title} {Physical origin of giant excitonic and magneto-optical
  responses in two-dimensional ferromagnetic insulators},}\ }\href@noop {}
  {\bibfield  {journal} {\bibinfo  {journal} {Nature Comm.}\ }\textbf {\bibinfo
  {volume} {10}},\ \bibinfo {pages} {2371} (\bibinfo {year}
  {2019})}\BibitemShut {NoStop}%
\bibitem [{\citenamefont {Chernikov}\ \emph {et~al.}(2014)\citenamefont
  {Chernikov}, \citenamefont {Berkelbach}, \citenamefont {Hill}, \citenamefont
  {Rigosi}, \citenamefont {Li}, \citenamefont {Aslan}, \citenamefont
  {Reichman}, \citenamefont {Hybertsen},\ and\ \citenamefont
  {Heinz}}]{Chernikov2014}%
  \BibitemOpen
  \bibfield  {author} {\bibinfo {author} {\bibfnamefont {A.}~\bibnamefont
  {Chernikov}}, \bibinfo {author} {\bibfnamefont {T.~C.}\ \bibnamefont
  {Berkelbach}}, \bibinfo {author} {\bibfnamefont {H.~M.}\ \bibnamefont
  {Hill}}, \bibinfo {author} {\bibfnamefont {A.}~\bibnamefont {Rigosi}},
  \bibinfo {author} {\bibfnamefont {Y.}~\bibnamefont {Li}}, \bibinfo {author}
  {\bibfnamefont {O.~B.}\ \bibnamefont {Aslan}}, \bibinfo {author}
  {\bibfnamefont {D.~R.}\ \bibnamefont {Reichman}}, \bibinfo {author}
  {\bibfnamefont {M.~S.}\ \bibnamefont {Hybertsen}}, \ and\ \bibinfo {author}
  {\bibfnamefont {T.~F.}\ \bibnamefont {Heinz}},\ }\bibfield  {title} {\enquote
  {\bibinfo {title} {Exciton binding energy and nonhydrogenic rydberg series in
  monolayer ${\mathrm{ws}}_{2}$},}\ }\href@noop {} {\bibfield  {journal}
  {\bibinfo  {journal} {Phys. Rev. Lett.}\ }\textbf {\bibinfo {volume} {113}},\
  \bibinfo {pages} {076802} (\bibinfo {year} {2014})}\BibitemShut {NoStop}%
\bibitem [{\citenamefont {Splendiani}\ \emph {et~al.}(2010)\citenamefont
  {Splendiani}, \citenamefont {Sun}, \citenamefont {Zhang}, \citenamefont {Li},
  \citenamefont {Kim}, \citenamefont {Chim}, \citenamefont {Galli},\ and\
  \citenamefont {Wang}}]{Splendiani2010}%
  \BibitemOpen
  \bibfield  {author} {\bibinfo {author} {\bibfnamefont {A.}~\bibnamefont
  {Splendiani}}, \bibinfo {author} {\bibfnamefont {L.}~\bibnamefont {Sun}},
  \bibinfo {author} {\bibfnamefont {Y.}~\bibnamefont {Zhang}}, \bibinfo
  {author} {\bibfnamefont {T.}~\bibnamefont {Li}}, \bibinfo {author}
  {\bibfnamefont {J.}~\bibnamefont {Kim}}, \bibinfo {author} {\bibfnamefont
  {C.-Y.}\ \bibnamefont {Chim}}, \bibinfo {author} {\bibfnamefont
  {G.}~\bibnamefont {Galli}}, \ and\ \bibinfo {author} {\bibfnamefont
  {F.}~\bibnamefont {Wang}},\ }\bibfield  {title} {\enquote {\bibinfo {title}
  {Emerging photoluminescence in monolayer ${\mathrm{mos}}_{2}$},}\ }\href@noop
  {} {\bibfield  {journal} {\bibinfo  {journal} {Nano Lett.}\ }\textbf
  {\bibinfo {volume} {10}},\ \bibinfo {pages} {1271} (\bibinfo {year}
  {2010})}\BibitemShut {NoStop}%
\bibitem [{\citenamefont {Steinleitner}\ \emph {et~al.}(2017)\citenamefont
  {Steinleitner}, \citenamefont {Merkl}, \citenamefont {Nagler}, \citenamefont
  {Mornhinweg}, \citenamefont {Sch\"{u}ller}, \citenamefont {Korn},
  \citenamefont {Chernikov},\ and\ \citenamefont {Huber}}]{Steinleitner2017}%
  \BibitemOpen
  \bibfield  {author} {\bibinfo {author} {\bibfnamefont {P.}~\bibnamefont
  {Steinleitner}}, \bibinfo {author} {\bibfnamefont {P.}~\bibnamefont {Merkl}},
  \bibinfo {author} {\bibfnamefont {P.}~\bibnamefont {Nagler}}, \bibinfo
  {author} {\bibfnamefont {J.}~\bibnamefont {Mornhinweg}}, \bibinfo {author}
  {\bibfnamefont {C.}~\bibnamefont {Sch\"{u}ller}}, \bibinfo {author}
  {\bibfnamefont {T.}~\bibnamefont {Korn}}, \bibinfo {author} {\bibfnamefont
  {A.}~\bibnamefont {Chernikov}}, \ and\ \bibinfo {author} {\bibfnamefont
  {R.}~\bibnamefont {Huber}},\ }\bibfield  {title} {\enquote {\bibinfo {title}
  {Direct observation of ultrafast exciton formation in a monolayer of
  ${\mathrm{wse}}_{2}$},}\ }\href@noop {} {\bibfield  {journal} {\bibinfo
  {journal} {Nano Lett.}\ }\textbf {\bibinfo {volume} {17}},\ \bibinfo {pages}
  {1455} (\bibinfo {year} {2017})}\BibitemShut {NoStop}%
\bibitem [{\citenamefont {Wang}\ \emph {et~al.}(2018)\citenamefont {Wang},
  \citenamefont {Chernikov}, \citenamefont {Glazov}, \citenamefont {Heinz},
  \citenamefont {Marie}, \citenamefont {Amand},\ and\ \citenamefont
  {Urbaszek}}]{Wang2018}%
  \BibitemOpen
  \bibfield  {author} {\bibinfo {author} {\bibfnamefont {G.}~\bibnamefont
  {Wang}}, \bibinfo {author} {\bibfnamefont {A.}~\bibnamefont {Chernikov}},
  \bibinfo {author} {\bibfnamefont {M.~M.}\ \bibnamefont {Glazov}}, \bibinfo
  {author} {\bibfnamefont {T.~F.}\ \bibnamefont {Heinz}}, \bibinfo {author}
  {\bibfnamefont {X.}~\bibnamefont {Marie}}, \bibinfo {author} {\bibfnamefont
  {T.}~\bibnamefont {Amand}}, \ and\ \bibinfo {author} {\bibfnamefont
  {B.}~\bibnamefont {Urbaszek}},\ }\bibfield  {title} {\enquote {\bibinfo
  {title} {Colloquium: Excitons in atomically thin transition metal
  dichalcogenides},}\ }\href@noop {} {\bibfield  {journal} {\bibinfo  {journal}
  {Rev. Mod. Phys.}\ }\textbf {\bibinfo {volume} {90}},\ \bibinfo {pages}
  {021001} (\bibinfo {year} {2018})}\BibitemShut {NoStop}%
\bibitem [{\citenamefont {Kudlis}\ \emph {et~al.}(2021)\citenamefont {Kudlis},
  \citenamefont {Iorsh},\ and\ \citenamefont {Shelykh}}]{PhysRevB.104.L020412}%
  \BibitemOpen
  \bibfield  {author} {\bibinfo {author} {\bibfnamefont {A.}~\bibnamefont
  {Kudlis}}, \bibinfo {author} {\bibfnamefont {I.}~\bibnamefont {Iorsh}}, \
  and\ \bibinfo {author} {\bibfnamefont {I.~A.}\ \bibnamefont {Shelykh}},\
  }\bibfield  {title} {\enquote {\bibinfo {title} {All-optical resonant
  magnetization switching in ${\mathrm{cri}}_{3}$ monolayers},}\ }\href
  {\doibase 10.1103/PhysRevB.104.L020412} {\bibfield  {journal} {\bibinfo
  {journal} {Phys. Rev. B}\ }\textbf {\bibinfo {volume} {104}},\ \bibinfo
  {pages} {L020412} (\bibinfo {year} {2021})}\BibitemShut {NoStop}%
\bibitem [{\citenamefont {Mortensen}\ \emph
  {et~al.}(2005{\natexlab{a}})\citenamefont {Mortensen}, \citenamefont
  {Hansen},\ and\ \citenamefont {Jacobsen}}]{Mortensen_2005}%
  \BibitemOpen
  \bibfield  {author} {\bibinfo {author} {\bibfnamefont {J.~J.}\ \bibnamefont
  {Mortensen}}, \bibinfo {author} {\bibfnamefont {L.~B.}\ \bibnamefont
  {Hansen}}, \ and\ \bibinfo {author} {\bibfnamefont {K.~W.}\ \bibnamefont
  {Jacobsen}},\ }\bibfield  {title} {\enquote {\bibinfo {title} {Real-space
  grid implementation of the projector augmented wave method},}\ }\href
  {\doibase 10.1103/physrevb.71.035109} {\bibfield  {journal} {\bibinfo
  {journal} {Physical Review B}\ }\textbf {\bibinfo {volume} {71}} (\bibinfo
  {year} {2005}{\natexlab{a}}),\ 10.1103/physrevb.71.035109}\BibitemShut
  {NoStop}%
\bibitem [{\citenamefont {Enkovaara}\ \emph {et~al.}(2010)\citenamefont
  {Enkovaara}, \citenamefont {Rostgaard}, \citenamefont {Mortensen},
  \citenamefont {Chen}, \citenamefont {Du{\l}ak}, \citenamefont {Ferrighi},
  \citenamefont {Gavnholt}, \citenamefont {Glinsvad}, \citenamefont {Haikola},
  \citenamefont {Hansen}, \citenamefont {Kristoffersen}, \citenamefont
  {Kuisma}, \citenamefont {Larsen}, \citenamefont {Lehtovaara}, \citenamefont
  {Ljungberg}, \citenamefont {Lopez-Acevedo}, \citenamefont {Moses},
  \citenamefont {Ojanen}, \citenamefont {Olsen}, \citenamefont {Petzold},
  \citenamefont {Romero}, \citenamefont {Stausholm-M{\o}ller}, \citenamefont
  {Strange}, \citenamefont {Tritsaris}, \citenamefont {Vanin}, \citenamefont
  {Walter}, \citenamefont {Hammer}, \citenamefont {Häkkinen}, \citenamefont
  {Madsen}, \citenamefont {Nieminen}, \citenamefont {N{\o}rskov}, \citenamefont
  {Puska}, \citenamefont {Rantala}, \citenamefont {Schi{\o}tz}, \citenamefont
  {Thygesen},\ and\ \citenamefont {Jacobsen}}]{Enkovaara_2010}%
  \BibitemOpen
  \bibfield  {author} {\bibinfo {author} {\bibfnamefont {J}~\bibnamefont
  {Enkovaara}}, \bibinfo {author} {\bibfnamefont {C}~\bibnamefont {Rostgaard}},
  \bibinfo {author} {\bibfnamefont {J~J}\ \bibnamefont {Mortensen}}, \bibinfo
  {author} {\bibfnamefont {J}~\bibnamefont {Chen}}, \bibinfo {author}
  {\bibfnamefont {M}~\bibnamefont {Du{\l}ak}}, \bibinfo {author} {\bibfnamefont
  {L}~\bibnamefont {Ferrighi}}, \bibinfo {author} {\bibfnamefont
  {J}~\bibnamefont {Gavnholt}}, \bibinfo {author} {\bibfnamefont
  {C}~\bibnamefont {Glinsvad}}, \bibinfo {author} {\bibfnamefont
  {V}~\bibnamefont {Haikola}}, \bibinfo {author} {\bibfnamefont {H~A}\
  \bibnamefont {Hansen}}, \bibinfo {author} {\bibfnamefont {H~H}\ \bibnamefont
  {Kristoffersen}}, \bibinfo {author} {\bibfnamefont {M}~\bibnamefont
  {Kuisma}}, \bibinfo {author} {\bibfnamefont {A~H}\ \bibnamefont {Larsen}},
  \bibinfo {author} {\bibfnamefont {L}~\bibnamefont {Lehtovaara}}, \bibinfo
  {author} {\bibfnamefont {M}~\bibnamefont {Ljungberg}}, \bibinfo {author}
  {\bibfnamefont {O}~\bibnamefont {Lopez-Acevedo}}, \bibinfo {author}
  {\bibfnamefont {P~G}\ \bibnamefont {Moses}}, \bibinfo {author} {\bibfnamefont
  {J}~\bibnamefont {Ojanen}}, \bibinfo {author} {\bibfnamefont {T}~\bibnamefont
  {Olsen}}, \bibinfo {author} {\bibfnamefont {V}~\bibnamefont {Petzold}},
  \bibinfo {author} {\bibfnamefont {N~A}\ \bibnamefont {Romero}}, \bibinfo
  {author} {\bibfnamefont {J}~\bibnamefont {Stausholm-M{\o}ller}}, \bibinfo
  {author} {\bibfnamefont {M}~\bibnamefont {Strange}}, \bibinfo {author}
  {\bibfnamefont {G~A}\ \bibnamefont {Tritsaris}}, \bibinfo {author}
  {\bibfnamefont {M}~\bibnamefont {Vanin}}, \bibinfo {author} {\bibfnamefont
  {M}~\bibnamefont {Walter}}, \bibinfo {author} {\bibfnamefont {B}~\bibnamefont
  {Hammer}}, \bibinfo {author} {\bibfnamefont {H}~\bibnamefont {Häkkinen}},
  \bibinfo {author} {\bibfnamefont {G~K~H}\ \bibnamefont {Madsen}}, \bibinfo
  {author} {\bibfnamefont {R~M}\ \bibnamefont {Nieminen}}, \bibinfo {author}
  {\bibfnamefont {J~K}\ \bibnamefont {N{\o}rskov}}, \bibinfo {author}
  {\bibfnamefont {M}~\bibnamefont {Puska}}, \bibinfo {author} {\bibfnamefont
  {T~T}\ \bibnamefont {Rantala}}, \bibinfo {author} {\bibfnamefont
  {J}~\bibnamefont {Schi{\o}tz}}, \bibinfo {author} {\bibfnamefont {K~S}\
  \bibnamefont {Thygesen}}, \ and\ \bibinfo {author} {\bibfnamefont {K~W}\
  \bibnamefont {Jacobsen}},\ }\bibfield  {title} {\enquote {\bibinfo {title}
  {Electronic structure calculations with {GPAW}: a real-space implementation
  of the projector augmented-wave method},}\ }\href {\doibase
  10.1088/0953-8984/22/25/253202} {\bibfield  {journal} {\bibinfo  {journal}
  {Journal of Physics: Condensed Matter}\ }\textbf {\bibinfo {volume} {22}},\
  \bibinfo {pages} {253202} (\bibinfo {year} {2010})}\BibitemShut {NoStop}%
\bibitem [{\citenamefont {Olsen}(2016)}]{PhysRevB.94.235106}%
  \BibitemOpen
  \bibfield  {author} {\bibinfo {author} {\bibfnamefont {Thomas}\ \bibnamefont
  {Olsen}},\ }\bibfield  {title} {\enquote {\bibinfo {title} {Designing
  in-plane heterostructures of quantum spin hall insulators from first
  principles: $1{\mathrm{t}}^{\ensuremath{'}}\ensuremath{-}{\mathrm{mos}}_{2}$
  with adsorbates},}\ }\href {\doibase 10.1103/PhysRevB.94.235106} {\bibfield
  {journal} {\bibinfo  {journal} {Phys. Rev. B}\ }\textbf {\bibinfo {volume}
  {94}},\ \bibinfo {pages} {235106} (\bibinfo {year} {2016})}\BibitemShut
  {NoStop}%
\bibitem [{\citenamefont {Rohlfing}\ and\ \citenamefont
  {Louie}(2000)}]{PhysRevB.62.4927}%
  \BibitemOpen
  \bibfield  {author} {\bibinfo {author} {\bibfnamefont {Michael}\ \bibnamefont
  {Rohlfing}}\ and\ \bibinfo {author} {\bibfnamefont {Steven~G.}\ \bibnamefont
  {Louie}},\ }\bibfield  {title} {\enquote {\bibinfo {title} {Electron-hole
  excitations and optical spectra from first principles},}\ }\href {\doibase
  10.1103/PhysRevB.62.4927} {\bibfield  {journal} {\bibinfo  {journal} {Phys.
  Rev. B}\ }\textbf {\bibinfo {volume} {62}},\ \bibinfo {pages} {4927--4944}
  (\bibinfo {year} {2000})}\BibitemShut {NoStop}%
\bibitem [{\citenamefont {Yan}\ \emph {et~al.}(2011)\citenamefont {Yan},
  \citenamefont {Mortensen}, \citenamefont {Jacobsen},\ and\ \citenamefont
  {Thygesen}}]{PhysRevB.83.245122}%
  \BibitemOpen
  \bibfield  {author} {\bibinfo {author} {\bibfnamefont {Jun}\ \bibnamefont
  {Yan}}, \bibinfo {author} {\bibfnamefont {Jens.~J.}\ \bibnamefont
  {Mortensen}}, \bibinfo {author} {\bibfnamefont {Karsten~W.}\ \bibnamefont
  {Jacobsen}}, \ and\ \bibinfo {author} {\bibfnamefont {Kristian~S.}\
  \bibnamefont {Thygesen}},\ }\bibfield  {title} {\enquote {\bibinfo {title}
  {Linear density response function in the projector augmented wave method:
  Applications to solids, surfaces, and interfaces},}\ }\href {\doibase
  10.1103/PhysRevB.83.245122} {\bibfield  {journal} {\bibinfo  {journal} {Phys.
  Rev. B}\ }\textbf {\bibinfo {volume} {83}},\ \bibinfo {pages} {245122}
  (\bibinfo {year} {2011})}\BibitemShut {NoStop}%
\bibitem [{\citenamefont {H\"user}\ \emph {et~al.}(2013)\citenamefont
  {H\"user}, \citenamefont {Olsen},\ and\ \citenamefont
  {Thygesen}}]{PhysRevB.88.245309}%
  \BibitemOpen
  \bibfield  {author} {\bibinfo {author} {\bibfnamefont {Falco}\ \bibnamefont
  {H\"user}}, \bibinfo {author} {\bibfnamefont {Thomas}\ \bibnamefont {Olsen}},
  \ and\ \bibinfo {author} {\bibfnamefont {Kristian~S.}\ \bibnamefont
  {Thygesen}},\ }\bibfield  {title} {\enquote {\bibinfo {title} {How dielectric
  screening in two-dimensional crystals affects the convergence of
  excited-state calculations: Monolayer mos${}_{2}$},}\ }\href {\doibase
  10.1103/PhysRevB.88.245309} {\bibfield  {journal} {\bibinfo  {journal} {Phys.
  Rev. B}\ }\textbf {\bibinfo {volume} {88}},\ \bibinfo {pages} {245309}
  (\bibinfo {year} {2013})}\BibitemShut {NoStop}%
\bibitem [{\citenamefont {Olsen}(2021)}]{PhysRevLett.127.166402}%
  \BibitemOpen
  \bibfield  {author} {\bibinfo {author} {\bibfnamefont {Thomas}\ \bibnamefont
  {Olsen}},\ }\bibfield  {title} {\enquote {\bibinfo {title} {Unified treatment
  of magnons and excitons in monolayer ${\mathrm{cri}}_{3}$ from many-body
  perturbation theory},}\ }\href {\doibase 10.1103/PhysRevLett.127.166402}
  {\bibfield  {journal} {\bibinfo  {journal} {Phys. Rev. Lett.}\ }\textbf
  {\bibinfo {volume} {127}},\ \bibinfo {pages} {166402} (\bibinfo {year}
  {2021})}\BibitemShut {NoStop}%
\bibitem [{\citenamefont {Ghosh}\ \emph {et~al.}(2020)\citenamefont {Ghosh},
  \citenamefont {Stoji{\'c}},\ and\ \citenamefont
  {Binggeli}}]{ghosh2020comment}%
  \BibitemOpen
  \bibfield  {author} {\bibinfo {author} {\bibfnamefont {Sukanya}\ \bibnamefont
  {Ghosh}}, \bibinfo {author} {\bibfnamefont {Nata{\v{s}}a}\ \bibnamefont
  {Stoji{\'c}}}, \ and\ \bibinfo {author} {\bibfnamefont {Nadia}\ \bibnamefont
  {Binggeli}},\ }\bibfield  {title} {\enquote {\bibinfo {title} {Comment on
  “magnetic skyrmions in atomic thin cri 3 monolayer”[appl. phys. lett.
  114, 232402 (2019)]},}\ }\href@noop {} {\bibfield  {journal} {\bibinfo
  {journal} {Applied Physics Letters}\ }\textbf {\bibinfo {volume} {116}},\
  \bibinfo {pages} {086101} (\bibinfo {year} {2020})}\BibitemShut {NoStop}%
\bibitem [{\citenamefont {Mentink}\ \emph {et~al.}(2010)\citenamefont
  {Mentink}, \citenamefont {Tretyakov}, \citenamefont {Fasolino}, \citenamefont
  {Katsnelson},\ and\ \citenamefont {Rasing}}]{mentink2010stable}%
  \BibitemOpen
  \bibfield  {author} {\bibinfo {author} {\bibfnamefont {JH}~\bibnamefont
  {Mentink}}, \bibinfo {author} {\bibfnamefont {MV}~\bibnamefont {Tretyakov}},
  \bibinfo {author} {\bibfnamefont {A}~\bibnamefont {Fasolino}}, \bibinfo
  {author} {\bibfnamefont {MI}~\bibnamefont {Katsnelson}}, \ and\ \bibinfo
  {author} {\bibfnamefont {Th}~\bibnamefont {Rasing}},\ }\bibfield  {title}
  {\enquote {\bibinfo {title} {Stable and fast semi-implicit integration of the
  stochastic landau--lifshitz equation},}\ }\href@noop {} {\bibfield  {journal}
  {\bibinfo  {journal} {Journal of Physics: Condensed Matter}\ }\textbf
  {\bibinfo {volume} {22}},\ \bibinfo {pages} {176001} (\bibinfo {year}
  {2010})}\BibitemShut {NoStop}%
\bibitem [{\citenamefont {Maciej}\ \emph {et~al.}(2022)\citenamefont {Maciej},
  \citenamefont {Shi}, \citenamefont {Mara}, \citenamefont {Paul},
  \citenamefont {Freddie}, \citenamefont {Santos},\ and\ \citenamefont
  {Hicken}}]{Dabrowski2022}%
  \BibitemOpen
  \bibfield  {author} {\bibinfo {author} {\bibfnamefont {Dacbrowski}\
  \bibnamefont {Maciej}}, \bibinfo {author} {\bibfnamefont {Guo}\ \bibnamefont
  {Shi}}, \bibinfo {author} {\bibfnamefont {Strungaru}\ \bibnamefont {Mara}},
  \bibinfo {author} {\bibfnamefont {S.~Keatley}\ \bibnamefont {Paul}}, \bibinfo
  {author} {\bibfnamefont {Withers}\ \bibnamefont {Freddie}}, \bibinfo {author}
  {\bibfnamefont {Elton J.~G.}\ \bibnamefont {Santos}}, \ and\ \bibinfo
  {author} {\bibfnamefont {Robert~J.}\ \bibnamefont {Hicken}},\ }\bibfield
  {title} {\enquote {\bibinfo {title} {All-optical control of spin in a 2d van
  der waals magnet},}\ }\href {\doibase 10.1038/s41467-022-33343-4} {\bibfield
  {journal} {\bibinfo  {journal} {Nature Communications}\ }\textbf {\bibinfo
  {volume} {13}} (\bibinfo {year} {2022}),\
  10.1038/s41467-022-33343-4}\BibitemShut {NoStop}%
\bibitem [{\citenamefont {Onida}\ \emph {et~al.}(2002)\citenamefont {Onida},
  \citenamefont {Reining},\ and\ \citenamefont {Rubio}}]{RevModPhys.74.601}%
  \BibitemOpen
  \bibfield  {author} {\bibinfo {author} {\bibfnamefont {Giovanni}\
  \bibnamefont {Onida}}, \bibinfo {author} {\bibfnamefont {Lucia}\ \bibnamefont
  {Reining}}, \ and\ \bibinfo {author} {\bibfnamefont {Angel}\ \bibnamefont
  {Rubio}},\ }\bibfield  {title} {\enquote {\bibinfo {title} {Electronic
  excitations: density-functional versus many-body green's-function
  approaches},}\ }\href {\doibase 10.1103/RevModPhys.74.601} {\bibfield
  {journal} {\bibinfo  {journal} {Rev. Mod. Phys.}\ }\textbf {\bibinfo {volume}
  {74}},\ \bibinfo {pages} {601--659} (\bibinfo {year} {2002})}\BibitemShut
  {NoStop}%
\bibitem [{\citenamefont {Rohlfing}\ and\ \citenamefont
  {Louie}(1998)}]{PhysRevLett.81.2312}%
  \BibitemOpen
  \bibfield  {author} {\bibinfo {author} {\bibfnamefont {Michael}\ \bibnamefont
  {Rohlfing}}\ and\ \bibinfo {author} {\bibfnamefont {Steven~G.}\ \bibnamefont
  {Louie}},\ }\bibfield  {title} {\enquote {\bibinfo {title} {Electron-hole
  excitations in semiconductors and insulators},}\ }\href {\doibase
  10.1103/PhysRevLett.81.2312} {\bibfield  {journal} {\bibinfo  {journal}
  {Phys. Rev. Lett.}\ }\textbf {\bibinfo {volume} {81}},\ \bibinfo {pages}
  {2312--2315} (\bibinfo {year} {1998})}\BibitemShut {NoStop}%
\bibitem [{\citenamefont {Deilmann}\ \emph {et~al.}(2020)\citenamefont
  {Deilmann}, \citenamefont {Kr\"uger},\ and\ \citenamefont
  {Rohlfing}}]{PhysRevLett.124.226402}%
  \BibitemOpen
  \bibfield  {author} {\bibinfo {author} {\bibfnamefont {Thorsten}\
  \bibnamefont {Deilmann}}, \bibinfo {author} {\bibfnamefont {Peter}\
  \bibnamefont {Kr\"uger}}, \ and\ \bibinfo {author} {\bibfnamefont {Michael}\
  \bibnamefont {Rohlfing}},\ }\bibfield  {title} {\enquote {\bibinfo {title}
  {Ab initio studies of exciton $g$ factors: Monolayer transition metal
  dichalcogenides in magnetic fields},}\ }\href {\doibase
  10.1103/PhysRevLett.124.226402} {\bibfield  {journal} {\bibinfo  {journal}
  {Phys. Rev. Lett.}\ }\textbf {\bibinfo {volume} {124}},\ \bibinfo {pages}
  {226402} (\bibinfo {year} {2020})}\BibitemShut {NoStop}%
\bibitem [{\citenamefont {Wo\ifmmode~\acute{z}\else \'{z}\fi{}niak}\ \emph
  {et~al.}(2020)\citenamefont {Wo\ifmmode~\acute{z}\else \'{z}\fi{}niak},
  \citenamefont {Faria~Junior}, \citenamefont {Seifert}, \citenamefont
  {Chaves},\ and\ \citenamefont {Kunstmann}}]{PhysRevB.101.235408}%
  \BibitemOpen
  \bibfield  {author} {\bibinfo {author} {\bibfnamefont {Tomasz}\ \bibnamefont
  {Wo\ifmmode~\acute{z}\else \'{z}\fi{}niak}}, \bibinfo {author} {\bibfnamefont
  {Paulo~E.}\ \bibnamefont {Faria~Junior}}, \bibinfo {author} {\bibfnamefont
  {Gotthard}\ \bibnamefont {Seifert}}, \bibinfo {author} {\bibfnamefont
  {Andrey}\ \bibnamefont {Chaves}}, \ and\ \bibinfo {author} {\bibfnamefont
  {Jens}\ \bibnamefont {Kunstmann}},\ }\bibfield  {title} {\enquote {\bibinfo
  {title} {Exciton $g$ factors of van der waals heterostructures from
  first-principles calculations},}\ }\href {\doibase
  10.1103/PhysRevB.101.235408} {\bibfield  {journal} {\bibinfo  {journal}
  {Phys. Rev. B}\ }\textbf {\bibinfo {volume} {101}},\ \bibinfo {pages}
  {235408} (\bibinfo {year} {2020})}\BibitemShut {NoStop}%
\bibitem [{\citenamefont {Lopez}\ \emph {et~al.}(2012)\citenamefont {Lopez},
  \citenamefont {Vanderbilt}, \citenamefont {Thonhauser},\ and\ \citenamefont
  {Souza}}]{PhysRevB.85.014435}%
  \BibitemOpen
  \bibfield  {author} {\bibinfo {author} {\bibfnamefont {M.~G.}\ \bibnamefont
  {Lopez}}, \bibinfo {author} {\bibfnamefont {David}\ \bibnamefont
  {Vanderbilt}}, \bibinfo {author} {\bibfnamefont {T.}~\bibnamefont
  {Thonhauser}}, \ and\ \bibinfo {author} {\bibfnamefont {Ivo}\ \bibnamefont
  {Souza}},\ }\bibfield  {title} {\enquote {\bibinfo {title} {Wannier-based
  calculation of the orbital magnetization in crystals},}\ }\href {\doibase
  10.1103/PhysRevB.85.014435} {\bibfield  {journal} {\bibinfo  {journal} {Phys.
  Rev. B}\ }\textbf {\bibinfo {volume} {85}},\ \bibinfo {pages} {014435}
  (\bibinfo {year} {2012})}\BibitemShut {NoStop}%
\bibitem [{\citenamefont {Thonhauser}\ \emph {et~al.}(2005)\citenamefont
  {Thonhauser}, \citenamefont {Ceresoli}, \citenamefont {Vanderbilt},\ and\
  \citenamefont {Resta}}]{PhysRevLett.95.137205}%
  \BibitemOpen
  \bibfield  {author} {\bibinfo {author} {\bibfnamefont {T.}~\bibnamefont
  {Thonhauser}}, \bibinfo {author} {\bibfnamefont {Davide}\ \bibnamefont
  {Ceresoli}}, \bibinfo {author} {\bibfnamefont {David}\ \bibnamefont
  {Vanderbilt}}, \ and\ \bibinfo {author} {\bibfnamefont {R.}~\bibnamefont
  {Resta}},\ }\bibfield  {title} {\enquote {\bibinfo {title} {Orbital
  magnetization in periodic insulators},}\ }\href {\doibase
  10.1103/PhysRevLett.95.137205} {\bibfield  {journal} {\bibinfo  {journal}
  {Phys. Rev. Lett.}\ }\textbf {\bibinfo {volume} {95}},\ \bibinfo {pages}
  {137205} (\bibinfo {year} {2005})}\BibitemShut {NoStop}%
\bibitem [{\citenamefont {Ceresoli}\ \emph {et~al.}(2006)\citenamefont
  {Ceresoli}, \citenamefont {Thonhauser}, \citenamefont {Vanderbilt},\ and\
  \citenamefont {Resta}}]{PhysRevB.74.024408}%
  \BibitemOpen
  \bibfield  {author} {\bibinfo {author} {\bibfnamefont {Davide}\ \bibnamefont
  {Ceresoli}}, \bibinfo {author} {\bibfnamefont {T.}~\bibnamefont
  {Thonhauser}}, \bibinfo {author} {\bibfnamefont {David}\ \bibnamefont
  {Vanderbilt}}, \ and\ \bibinfo {author} {\bibfnamefont {R.}~\bibnamefont
  {Resta}},\ }\bibfield  {title} {\enquote {\bibinfo {title} {Orbital
  magnetization in crystalline solids: Multi-band insulators, chern insulators,
  and metals},}\ }\href {\doibase 10.1103/PhysRevB.74.024408} {\bibfield
  {journal} {\bibinfo  {journal} {Phys. Rev. B}\ }\textbf {\bibinfo {volume}
  {74}},\ \bibinfo {pages} {024408} (\bibinfo {year} {2006})}\BibitemShut
  {NoStop}%
\bibitem [{\citenamefont {Mortensen}\ \emph
  {et~al.}(2005{\natexlab{b}})\citenamefont {Mortensen}, \citenamefont
  {Hansen},\ and\ \citenamefont {Jacobsen}}]{PhysRevB.71.035109}%
  \BibitemOpen
  \bibfield  {author} {\bibinfo {author} {\bibfnamefont {J.~J.}\ \bibnamefont
  {Mortensen}}, \bibinfo {author} {\bibfnamefont {L.~B.}\ \bibnamefont
  {Hansen}}, \ and\ \bibinfo {author} {\bibfnamefont {K.~W.}\ \bibnamefont
  {Jacobsen}},\ }\bibfield  {title} {\enquote {\bibinfo {title} {Real-space
  grid implementation of the projector augmented wave method},}\ }\href
  {\doibase 10.1103/PhysRevB.71.035109} {\bibfield  {journal} {\bibinfo
  {journal} {Phys. Rev. B}\ }\textbf {\bibinfo {volume} {71}},\ \bibinfo
  {pages} {035109} (\bibinfo {year} {2005}{\natexlab{b}})}\BibitemShut
  {NoStop}%
\bibitem [{\citenamefont {Bl\"ochl}(1994)}]{PhysRevB.50.17953}%
  \BibitemOpen
  \bibfield  {author} {\bibinfo {author} {\bibfnamefont {P.~E.}\ \bibnamefont
  {Bl\"ochl}},\ }\bibfield  {title} {\enquote {\bibinfo {title} {Projector
  augmented-wave method},}\ }\href {\doibase 10.1103/PhysRevB.50.17953}
  {\bibfield  {journal} {\bibinfo  {journal} {Phys. Rev. B}\ }\textbf {\bibinfo
  {volume} {50}},\ \bibinfo {pages} {17953--17979} (\bibinfo {year}
  {1994})}\BibitemShut {NoStop}%
\end{thebibliography}
%

\end{document}